\documentclass[reprint,aps,prb]{revtex4-2}

\usepackage[normalem]{ulem}
\usepackage{chemformula} % Formula subscripts using \ch{}
\usepackage[T1]{fontenc} % Use modern font encodings
\usepackage{lmodern}
\usepackage{braket}
\usepackage{amsmath}	
\usepackage{amssymb}
\usepackage{appendix}
\usepackage{hyperref}
\usepackage{epstopdf}
\usepackage{bm}
\usepackage{dcolumn}% Align table columns on decimal point
\usepackage[version=4]{mhchem}

\usepackage{graphicx}
\usepackage[percent]{overpic} 
\usepackage{color}
\usepackage{transparent} 
\usepackage[caption=false]{subfig}
\begin{document} 
	
	\title{Spin Wave Driven Skyrmions in a Bipartite Antiferromagnetic Lattice}
	
	\author{Michael Lau$^{1,2}$}
	\author{Wolfgang H{\"a}usler$^3$}
	\author{Michael Thorwart$^{1,2}$}
	\affiliation{$^1$I. Institut f{\"u}r Theoretische Physik, Universit{\"a}t Hamburg, Notkestra\ss{}e 9, 22607 Hamburg, Germany \\
		$^2$The Hamburg Centre for Ultrafast Imaging, Universit{\"a}t Hamburg, Luruper Chaussee 149, 22761 Hamburg, Germany\\
		$^3$ Institut f{\"u}r Physik, Universit{\"a}t Augsburg, Universit{\"a}tsstra\ss{}e 1, 86135 Augsburg, Germany}
	
	\begin{abstract}
	We show that a Skyrmion in a classical bipartite antiferromagnetic lattice can be spatially displaced in a controlled manner by externally applied spin waves.  We reveal the relation between the Skyrmion motion and the spin wave properties. To this end, we derive a classical spin wave formalism  which is tailored to the antiferromagnetic two-dimensional square lattice. 
	The antiferromagnetic spin waves can be classified into two types with respect to their polarization,  with two modes each. The circularly polarized spin waves oscillate with different amplitudes in the respective sublattices and induce a Skyrmion Hall effect. The two modes are symmetric under sublattices exchange and determine the overall sign of the Hall angle. For  linearly polarized spin waves, the two sublattices oscillate elliptically in opposite direction, however, with the same amplitude. These accelerate the Skyrmion solely into their own propagation direction. The two modes are symmetric under component $x$-$y$ exchange and  impact Bloch- or N{\'e}el Skyrmions differently. Our results indicate possible technical applications of spin-wave driven Skyrmion motion. As one example we propose a racetrack where spin waves pump Skyrmions along the track in antiferromagnets.
	\end{abstract}
	
	\maketitle
	\section{Introduction}
	\label{sec:Intro}
	
	Magnetic Skyrmions are vortex-like, non-collinear two-dimensional configurations of magnetic moments and are presently in the focus of attention as a research topic, not at least because of their promising technical applications as possible information carriers. Most commonly studied up to present are ferromagnetic Skyrmions, because they are relatively easy to detect in bulk solid-state materials or magnetic thin films. Their net magnetization also allows for a rather straightforward external manipulation by magnetic fields or spin-polarized electric currents. A comprehensive review of their topological and dynamical properties can be found in Ref.\ \cite{nagaosa_overview}. Ferromagnetic Skyrmions have initially been found in MnSi \cite{Sk1} and in FeCoSi \cite{Sk2} in Skyrmion lattices. In 2013, individual ultrasmall ferromagnetic Skyrmions in a PdFe bilayer grown on a (111) iridium surface have been reported \cite{Sk3} where even their internal magnetic structure could be resolved  \cite{sk_fm_exp_wiesendanger}. Use of Skyrmions as information carriers (bits) on a racetrack is tempting because they can be moved in a controlled manner by low-density electric currents \cite{fm_sk_racetrack}.
	
	 However, ferromagnetic Skyrmions are affected by the Skyrmion Hall Effect (SHE), which  deflects the skyrmion from its course along the racetrack. To counter this undesired effect, the idea came up to use antiferromagnetic (AFM) Skyrmions \cite{bogdanov_afm,zhang_afm_sk}. Naively, they would be expected to show no SHE due to the two ferromagnetic sublattices of opposite magnetization which form the antiferromagnet (in a semiclassical picture) and which compensate to zero net magnetization. Indeed, it was shown theoretically that Skyrmions in AFM materials can be stabilized \cite{bogdanov_afm}, created, and moved by electric currents, showing no SHE \cite{zhang_afm_sk, jin2016dynamics}. In synthetic AFMs, the controlled generation of Skyrmions using thermal effects has recently been reported  \cite{AFMSkExp22}. Likewise,  stable Skyrmions in synthetic AFMs  have been realized at room temperature \cite{legrand2020room} and also in Ferrimagnets \cite{caretta2018fast}. Yet, the realization of Skyrmions (or Antiskyrmions) in bipartite AFM lattices remains elusive up to present. 
	
	Ferromagnetic Skyrmions can be created \cite{StierPRL17, everschorsitte17} by spin-polarized, in-plane electric currents, or, even well controlled \cite{SieglPRB22}, by rotating the magnetization at the edge of a stripe \cite{SchaefferSciRep20}. Electric currents also allow to move ferromagnetic Skyrmions \cite{fm_sk_racetrack} and steer their Hall angle by electric gate voltages \cite{PlettenbergPRL20}. Another option is to use spin waves (magnons) in ferromagnets to drive ferromagnetic Skyrmions \cite{sw_iwasaki, schuette, schroeter_garst}. This approach has the advantage that no electric charges flow. Spin waves in ferromagnets can be created, for instance, by a pulse of a spin-polarized current through the magnetic tip of an atomic force microscope. Such a device would act as a point-like source of magnons. Once the magnons hit a ferromagnetic Skyrmion, the latter can be moved while the spin wave experiences scattering and, in general, deflection. 
	
	In this work, we investigate the possibility of driving AFM Skyrmions by externally applied spin waves traveling through the AFM host and interacting with a Skyrmion. We consider a basic model of a two-dimensional square lattice of classical magnetic moments. This model is well-established for describing ferromagnetic systems and we use it for the bipartite AFM case of two ferromagnetic sublattices of opposite magnetization in a checkerboard arrangement. This approach is generic to describe Skyrmions, see, e.g., Ref.\ \cite{zhang_afm_sk}. In particular, we aim to illuminate the relationship between spin waves properties and the resulting Skyrmion motion. To this end, we start with an analysis of the spin wave formalism. Although a theoretical investigation of classical AFM  spin waves is not new \cite{kefferkittel,dos2020modeling}, we introduce here the formalism and the notation as a basis of a further analysis of the specific  spin-wave-driven Skyrmion motion by linearizing the equation of motion in Sec.\ \ref{sec:spinwave}. The formalism is tailored to the model we use and is derived from the Landau-Lifshitz-Gilbert equation of motion. We obtain explicit analytical expressions for the spin wave properties relevant for the present situation, which can directly be compared to the numerical simulations of the full model dynamics. Especially, we are able to separate the AFM spin wave into components on their respective sublattices with corresponding amplitudes. This turns out as important regarding Skyrmion driving. The fact that AFM spin waves can move Skyrmions has been studied in Ref.\ \cite{daniels_sk}, however with an ansatz different from that used in the present work, and with a different focus. 
	
	We consider a bipartite single layer AFM on a discrete lattice for both an analytic description of the spin waves and the numerical simulations. Furthermore, we use the linearized equations of motion to describe the spin waves and consider and simulate them to be injected at one  edge of the  lattice. By this, we are able to control  the spin wave properties such as their  amplitude and frequency. Here, we highlight the importance of the sublattice spin wave amplitudes and their symmetries for the actual Skyrmion motion.  In Sec.\ \ref{sec:sk-sw}, we determine the motion of the AFM  Skyrmions under the influence of various spin waves. In section \ref{sec:damping} we also investigate the case of non-vanishing Gilbert damping $\alpha \neq 0$ present in the Landau-Lifshitz-Gilbert equation of motion. In general, we find a much stronger impact of magnons on the Skyrmion dynamics in the AFM case as compared to the ferromagnetic case. On the one hand, this can lead to more strongly disturbed Skyrmion motion, especially at elevated temperatures. On the other hand, spin waves are consequently much more efficient to steer a guided motion of the Skyrmion, which would make magnon-driven AFM Skyrmions even more interesting since, unlike the case of ferromagnets \cite{daniels_sk}, they evade the SHE. Finally, in Sec.\ \ref{sec:racetrack} we apply our findings to propose a concept of how an AFM Skyrmion racetrack could be designed using realistically damped spin waves.
	
	\section{Model}
	\label{sec:Model}
	We consider, as a generic model, a two-dimensional square lattice with classical magnetic moments  $\bm{M}_i$ at site $i$ assumed to be magnetized to saturation. The vectors representing the magnetic moment in this model are scaled such that they have unit length $|\bm{M}_i|/M_S = 1$. The saturation $M_S$ is absorbed in the system parameters in the Hamiltonian 
	\begin{equation}
		\label{eq:Hamiltonian_general}
		H = -\frac{J}{2} \sum_{\braket{i,j}} \bm{M}_i \bm{M}_j - D \sum_{\braket{i,j}} \bm{d}_{ij} \left[ \bm{M}_i \times \bm{M}_j \right] - K \sum_i \left(M_i^z\right)^2,
	\end{equation}
	where $\sum_{\braket{i,j}}$ is a summation over nearest neighbors. The first term describes the magnetic exchange interaction of strength $J$. 
	The Dzyaloshinsky-Moriya interaction (DMI, second term) is given here in general form. Originally, it roots in a broken symmetry of the crystal lattice \cite{moriya}. We include it as a phenomenological term, where $D$ is a constant and the vector $\bm{d}_{ij}$ defines whether we use bulk DMI $\bm{d}_{ij} = \bm{r}_{ij}$ or interfacial DMI $\bm{d}_{ij} = \bm{r}_{ij} \times \bm{z}$. Here, $\bm{r}_{ij}$ is the distance vector connecting site $i$ with site $j$, while $\bm{z}$ is the unit vector in $z$-direction, perpendicular to the $x$-$y$-plane of the lattice. The last term in Eq.\ (\ref{eq:Hamiltonian_general}) describes the magnetocrystalline anisotropy in the $z$-direction, of strength $K$.

\begin{figure}[t!]
		\subfloat{\includegraphics[width=0.5\textwidth]{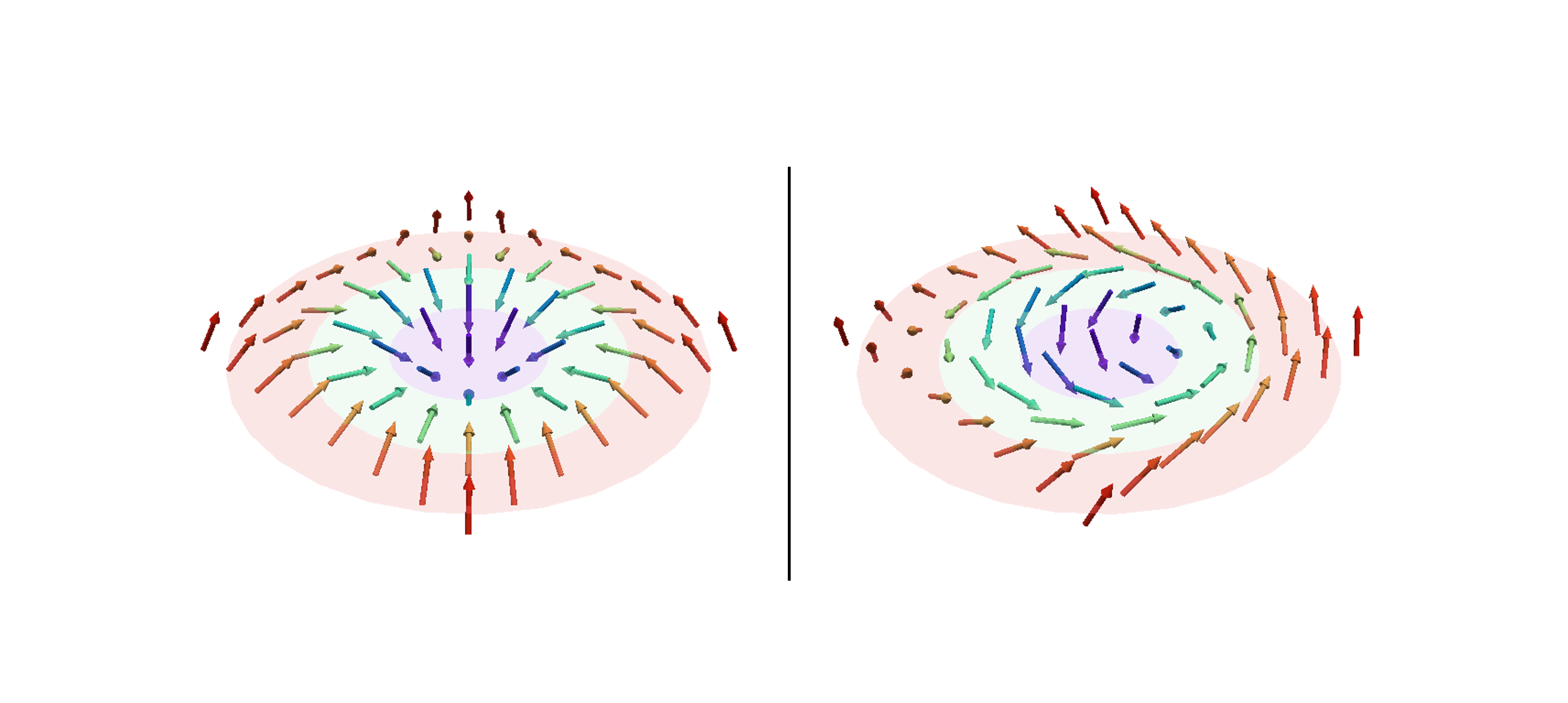}}
		\caption{(color online) Illustration of the magnetization profiles of a N{\'e}el Skyrmion (left) and a Bloch Skyrmion (right).
			\label{fig:image_bloch_neel}}
	\end{figure}

	In the ferromagnetic case, with the appropriate choice of parameters $D/J$ and $K/J$, this Hamiltonian stabilizes a Skyrmion \cite{sk_size}. Here, we consider both types: Bloch and N{\'e}el Skyrmions. Bloch Skyrmions are stabilized by bulk DMI, which is typically found in bulk materials with inversion symmetry breaking, while N{\'e}el Skyrmions are stabilized by the interfacial DMI of multilayers (see Ref.\ \cite{everschorsitte17}). Both types are illustrated in Fig.\ \ref{fig:image_bloch_neel}

	In the case of a bipartite antiferromagnet, we assume that the Hamiltonian in Eq.\ \eqref{eq:Hamiltonian_general} still holds. Then, a change of the parameters $J \rightarrow -J$ and $D \rightarrow -D$ stabilizes an AFM Skyrmion with the very same structure (size, chirality, etc.) \cite{footnote0}. For AFM Skyrmions, it is convenient to study them in the sublattice picture. Since neighboring moments try to align anti-parallel, the lattice becomes checkerboard-like, with  the 'white' fields denoted as sublattice A and the 'black' fields as sublattice B. We define a consistent notation for the magnetic moments depending on which sublattice they belong to, according to 
	\begin{equation}
		\label{eq:notation_sl}
		\bm{M}_i = \begin{cases}
			\bm{a}_i & \text{if } i \in \text{ sublattice A}, \\
			\bm{b}_i & \text{if } i \in \text{ sublattice B}.\\
		\end{cases}
	\end{equation}
	
	\section{Dynamics}
	The dynamics in our work is studied by the Landau-Lifshitz-Gilbert (LLG) equation \cite{gilbert}
	\begin{equation}
		\label{eq:LLG}
		\frac{\partial}{\partial t} \bm{M}_i = -\bm{M}_i \times \bm{H}_i^\text{eff} + \alpha \bm{M}_i \times \frac{\partial}{\partial t} \bm{M}_i.
	\end{equation}
	It describes the damped precession of a magnetic moment $\bm{M}_i$ around a local effective magnetic field $\bm{H}^\text{eff}_i$. We set the gyrocoupling constant $\gamma$ in such a way that the unit of time in our simulations is $t_0 = 1/J$. The effective magnetic field follows from the Hamiltonian as $\bm{H}_i^\text{eff} = - \partial H / \partial \bm{M}_i$ and, thus, from Eq.\ \eqref{eq:Hamiltonian_general} through
	\begin{equation}
		\label{eq:Heff_conc_fm}
		\bm{H}_i^\text{eff} = J \sum_{\bm{r}} \bm{M}_{i + \bm{r}} + 2D \sum_{\bm{r}} \left(\bm{M}_{i + \bm{r}} \times \bm{d}_{i, i + \bm{r}} \right) + 2K M_i^z \bm{z},
	\end{equation}
	where $\bm{r} \in \{\pm \bm{x}; \pm \bm{y}\}$ is the spatial vector pointing to the nearest neighbor.   In the sublattice picture of a semiclassical antiferromagnet, we have to consider two LLG equations of the form of Eq.\ \eqref{eq:LLG}, one for each sublattice corresponding to the cases in Eq.\ \eqref{eq:notation_sl}. Therefore, each sublattice has its own effective field $\bm{H}_{i,A}^\text{eff}$ (e.g.\ for sublattice A). Since the neighboring moments of a spin in sublattice A belong to sublattice B and vice versa, the effective field consists of moments from both sublattices and follows as 
	\begin{equation}
		\label{eq:heff_sl}
		\bm{H}_{i,A}^\text{eff} = J \sum_{\bm{r}} \bm{b}_{i + \bm{r}} + 2D \sum_{\bm{r}} \left(\bm{b}_{i + \bm{r}} \times \bm{d}_{i, i + \bm{r}} \right) + 2K a_i^z \bm{z},
	\end{equation} 
	and similar for $\bm{H}_{i,B}^\text{eff}$. This enables us to consider the antiferromagnet effectively as two separate ferromagnetic sublattices coupled by their respective effective fields $\bm{H}_{A/B}^\text{eff}$ \cite{BarTre2016}.
	
	\section{Spin Waves}
	\label{sec:spinwave}
	\subsection{Formalism}
	\label{sec:spinwave1}
	Before studying the spin wave - Skyrmion interaction, we first address global properties of the spin waves in the antiferromagnet in the absence of Gilbert damping, $\alpha=0$. Finite damping will be addressed in Sec.\ \ref{sec:damping}. To obtain results for harmonic spin waves, we start out with the linearized dynamics. The time evolution of a magnetic moment is approximated close to a stationary state according to $\bm{M}(t) \approx \bm{M}_0 + \delta \bm{M}(t)$. Inserting this decomposition in the LLG, Eq.\ \eqref{eq:LLG}, and linearizing we obtain the equation of motion  
	\begin{equation}
		\label{eq:lin_eom_nondamped}
		\delta \dot{\bm{M}}_i(t) = -\delta \bm{M}_i (t) \times \bm{H}_i^\text{eff,0} - \bm{M}_i^0 \times \delta \bm{H}_i^\text{eff}(t),
	\end{equation}    
	where the terms $\bm{H}_i^\text{eff,0}$ and $\delta \bm{H}_i^\text{eff}(t)$ are the effective fields given in  Eqs.\ \eqref{eq:Heff_conc_fm} and \eqref{eq:heff_sl} with $\bm{M}_i^0$ replaced by $\delta \bm{M}_i (t)$. Since we want to study spin waves around the classical AFM ground state, we assume, $\bm{a}_i^0 = \bm{z}$ and $\bm{b}_i^0 = - \bm{z}$ for each lattice site $i$, according to the notation of Eq.\ \eqref{eq:notation_sl}. Note that $\delta \bm{a}_i \cdot \bm{a}_i^0 = 0$ due to normalization. Using this we see that spin waves around the (anti-) ferromagnetic ground state are {\em independent} of the DMI: According to Eqs.\ \eqref{eq:lin_eom_nondamped} and \eqref{eq:Heff_conc_fm} two terms depend on the DMI. The first one is $\bm{H}_i^\text{eff,0}$ whose DMI-part $2D \sum_{\bm{r}} \left(\bm{M}^0_{i + \bm{r}} \times \bm{d}_{i, i + \bm{r}} \right)$ vanishes because the ground state $\bm{M}_i^0 = \bm{z}$ is independent of the lattice site. The other term is of the kind $\bm{M}_0 \times \left( \delta \bm{M} \times \bm{d} \right)$. This term vanishes because both $\delta \bm{M}$ and $\bm{d}$ are in the $xy$-plane and their cross product is thus parallel to the $z$ axis so that $\bm{M}_0 \times \bm{z}=0$. However, one should keep in mind that the spin waves are independent of the DMI only because we consider harmonic spin waves and a DMI perpendicular to the classical AFM ground state as used here. Without linearization or in the case of an in-plane antiferromagnet the DMI would affect spin waves \cite{gitgeatpong2017nonreciprocal} which, however, is not the case in the present work.	
	As already pointed out, in the sublattice picture we have two LLG equations and therefore obtain a system of two coupled linearized equations of motion for the AFM spin waves in the form 
	\begin{align}
		\label{eq:afm_lin_eom_i}
		\delta \dot{\bm{a}}_i &= \delta \bm{a}_i \times \left(J \sum_{\bm{r}} \bm{z} - 2K \bm{z} \right) - \bm{z} \times J \sum_{\bm{r}} \delta \bm{b}_{i + \bm{r}} \, , \\
		\label{eq:afm_lin_eom_ii}
		\delta \dot{\bm{b}}_i &= -\delta \bm{b}_i \times \left(J \sum_{\bm{r}} \bm{z} - 2K \bm{z} \right) + \bm{z} \times J \sum_{\bm{r}} \delta \bm{a}_{i + \bm{r}},
	\end{align}
	where we omit now the DMI. As an ansatz, we consider sublattice spin waves of the form $\delta \bm{a}_i = a (\cos(\bm{r_i} \bm{k} - \omega t), \sin(\bm{r_i} \bm{k} - \omega t), 0)^T$ and $\delta \bm{b}_i = b (\cos(\bm{r_i} \bm{k} - \omega t), \sin(\bm{r_i} \bm{k} - \omega t), 0)^T$ with the same wave vector $\bm{k}$ but different amplitudes $a$ and $b$. Using the trigonometric equalities  $\cos((\bm{r_i}+\bm{r}) \bm{k}) + \cos((\bm{r_i}-\bm{r}) \bm{k}) = 2 \cos(\bm{r_i} \bm{k}) \cos(\bm{r} \bm{k})$ and  $\sin((\bm{r_i}+\bm{r}) \bm{k}) + \sin((\bm{r_i}-\bm{r}) \bm{k}) = 2 \sin(\bm{r_i} \bm{k}) \cos(\bm{r} \bm{k})$, where $\bm{r}$ is the relative position between two neighboring sites, we arrive at the equations of motion
	\begin{align}
		\label{eq:afm_lin_eom_new}
		\delta \dot{\bm{a}}_i &= -\delta \bm{a}_i \times \rho \bm{z} - C_{\bm{k}} \delta \bm{b}_{i} \times \bm{z} \, ,\\
		\label{eq:afm_lin_eom_new_ii}
		\delta \dot{\bm{b}}_i &= \delta \bm{b}_i \times \rho \bm{z} + C_{\bm{k}} \delta \bm{a}_{i} \times \bm{z}.
	\end{align}
	Here, $\rho = 2K-4J$ which coincides with the homogeneous energy density $\rho = -2E_0/N$ of the classical AFM ground state, and
	\begin{equation}
		\label{eq:Ck}
		C_{\bm{k}} = -2 J [\cos(k_x d) + \cos(k_y d)]
	\end{equation}
	depends on the wave vector $\bm{k}=(k_x,k_y)$ and the lattice constant $d$. 
	With the previously mentioned assumption that $\delta \bm{a}$ and $\delta \bm{b}$ are each circularly oscillating waves, i.e.,\ the amplitudes are the same for the $x$ and $y$ component, it is possible to reduce the vector equation of motion of Eqs.\ \eqref{eq:afm_lin_eom_new} and \eqref{eq:afm_lin_eom_new_ii} to a system of coupled, linear differential equations regarding their amplitudes $a$ and $b$. In terms of the vector $(a,b)^T$ this system becomes a simple eigenvalue problem,
	\begin{equation}
		\label{eq:matrix_vec}
		\omega 
		\begin{pmatrix}
			a \\
			b
		\end{pmatrix} = \begin{pmatrix}
			-\rho & -C_{\bm{k}} \\
			C_{\bm{k}} & \rho
		\end{pmatrix} \begin{pmatrix}
			a \\
			b
		\end{pmatrix}.
	\end{equation} 
	The eigenvalues 
	\begin{equation}
		\label{eq:disp_rel}
		\omega(\bm{k}) = \pm \sqrt{\rho^2 - 4J^2 \left[ \cos(k_x d) + \cos(k_y d) \right]^2}
	\end{equation}
	 of the $2\times 2$-matrix come in pairs, are symmetric around zero and represent the dispersion relation. As spins oscillate circularly, we denote the corresponding waves {\em circularly polarized} with two modes for the two eigenvalues \cite{kefferkittel}. The fact that there arise two modes is a particular feature of an antiferromagnet; there is only one mode for a ferromagnet. Following Ref.\ \cite{daniels_sw} we call the $+\omega$-mode `left-handed', and the $-\omega$-mode `right-handed'.
	
	A particularly interesting quantity is the normalized eigenvector $\bm{\nu}$ associated with the eigenvalue $\omega$. Its components describe how strongly spins on the different sublattices oscillate. The eigenvectors obey the symmetries $\bm{\nu}(+\omega) = (\nu_1, \nu_2)$ and $\bm{\nu}(-\omega) = (-\nu_2, -\nu_1)$, depending on the eigenvalues, with the explicit expressions 
	\begin{equation}
		\label{eq:eig_vec_exp}
		\begin{aligned}
			\nu_1 &= \frac{-C_{\bm{k}}}{\sqrt{2 \rho \left(\rho + \sqrt{\rho^2 - C_{\bm{k}}^2}\right)}}\, , \\
			\nu_2 &= \frac{C_{\bm{k}}}{\sqrt{2 \rho \left(\rho - \sqrt{\rho^2 - C_{\bm{k}}^2}\right)}}\, .
		\end{aligned}
	\end{equation}
	These two quantities can, as components of the eigenvector, be interpreted as the normalized amplitude of the spin oscillation on the respective sublattice. Since $\nu_1$ and $\nu_2$ are not equal and both depend on $\bm{k}$, it means that the two sublattices oscillate not only in opposite direction but also with different amplitudes depending on the wave vector. It is noticeable that the right- and left handed spin waves are anti-symmetric under sublattice exchange.

	\subsection{Superposition of Spin Waves}
	\label{sec:spinwave2}
	Clearly, left- and right-handed circularly polarized spin waves can be superposed. In the following, we examine two kinds of superposition, namely the symmetric ($+$) and the anti-symmetric ($-$) sum of both parts. 
		
	In the derivation for the circularly polarized spin waves above, we have assumed the form, e.g., for sublattice A, $\delta \bm{a}_i = a (\cos(\bm{r_i} \bm{k} - \omega t), \sin(\bm{r_i} \bm{k} - \omega t), 0)^T$. This leads to the two solutions for $\omega$ given in Eq.\ \eqref{eq:disp_rel} and to amplitudes (eigenvectors) $\nu_1$ for $+\omega$ and $-\nu_2$ for $-\omega$. A different spin wave $\delta \bm{a}^{\prime}= a (\cos(\bm{r_i} \bm{k} - \omega t), -\sin(\bm{r_i} \bm{k} - \omega t), 0)^T$ of opposite circulation, also solves the equations of motion Eq.\ \eqref{eq:afm_lin_eom_new} and \eqref{eq:afm_lin_eom_new_ii} with the same two solutions. For such a spin wave, also the  amplitudes (eigenvectors) are exchanged, i.e., $-\nu_2$ for $+\omega$ and $\nu_1$ for $-\omega$. This is similar for the B sublattice. Due to the exchange of the  sublattice oscillation  amplitude, the superposition of the two differently oscillating sublattice waves does not vanish and   elliptically oscillating waves emerge, according to 
	\begin{align}
		\label{eq:sw_superpos}
		\delta \bm{a}(\omega) \pm \delta \bm{a}^{\prime}(\omega) &= \begin{pmatrix}
			[\nu_1 \mp \nu_2] \cos(\bm{r_i} \bm{k} - \omega t) \\
			[\nu_1 \pm \nu_2] \sin(\bm{r_i} \bm{k} - \omega t)
		\end{pmatrix}, \\
		\delta \bm{b}(\omega) \pm \delta \bm{b}^{\prime}(\omega) &= \begin{pmatrix}
			[\nu_2 \mp \nu_1] \cos(\bm{r_i} \bm{k} - \omega t) \\
			[\nu_2 \pm \nu_1] \sin(\bm{r_i} \bm{k} - \omega t)
	\end{pmatrix}.
	\end{align}
	They show some peculiarities. The respective amplitudes in $x$ and $y$ direction are the same for both sublattices A and B. While both sublattices oscillate with the same amplitudes and the same frequency, they oscillate in different directions. It is convenient to define a four-component vector $\psi_{\pm} = (a_x, b_x, a_y, b_y)$, which contains the amplitudes and its index denotes whether the superposition of the wave is symmetric ($+$) or anti-symmetric ($-$), i.e.,
	\begin{equation}
		\label{eq:eig_vec_lin}
		\psi_{+} = \begin{pmatrix}
			\nu_1 -\nu_2 \\
			-(\nu_1 -\nu_2)\\
			\nu_1 +\nu_2\\
			\nu_1 +\nu_2
		\end{pmatrix} \text{  and  }
		\psi_{-} = \begin{pmatrix}
			\nu_1 + \nu_2 \\
			\nu_1 + \nu_2\\
			\nu_1 -\nu_2\\
			-(\nu_1 -\nu_2)
		\end{pmatrix}.
	\end{equation}
	Only for this special case, we recover the known results of AFM magnons in terms of the staggered magnetization $\bm{l} = (\bm{a} - \bm{b})/2$ and the total magnetization $\bm{m} = (\bm{a} + \bm{b})/2$, see, e.g., Ref.\ \cite{bogdanov1989contribution}. One can easily see that for $\psi_{+}$, the staggered magnetization, exclusively oscillates in $x$-direction (and the magnetization exclusively in $y$-direction). In analogy to electromagnetic waves, as well as to Ref.\ \cite{daniels_sk}, we thus denote $\psi_{+}$ as \textit{linearly polarized spin waves} in the $x$-direction. For the same reason, we denote $\psi_{-}$ as {\em linearly polarized spin waves} in the $y$-direction. Nevertheless, we continue using the sublattice picture. In view of Eq.\ \eqref{eq:eig_vec_lin} and ignoring the sign, it is noticeable that for the linearly polarized spin waves the sublattices behave similarly, i.e., $|a_x|=|b_x|$ as well as $|a_y|=|b_y|$. Additionally, the spin waves do not oscillate circularly, but elliptically. Therefore, we may focus on the difference of the components. It turns out that the sign of the superposition creates a component symmetry $x \leftrightarrow y$. 
	The $x$-polarized spin wave oscillates with the amplitude $|\nu_1 - \nu_2|$ in the $x$-component and with $|\nu_1 + \nu_2|$ in the $y$-component, and vice versa for the $y$-polarized spin wave. 
	
	\section{Numerical Simulations of the Spin Waves}
	\label{sec:num_sw}
\begin{figure}[t!]
		\subfloat{\includegraphics[width=0.48\textwidth]{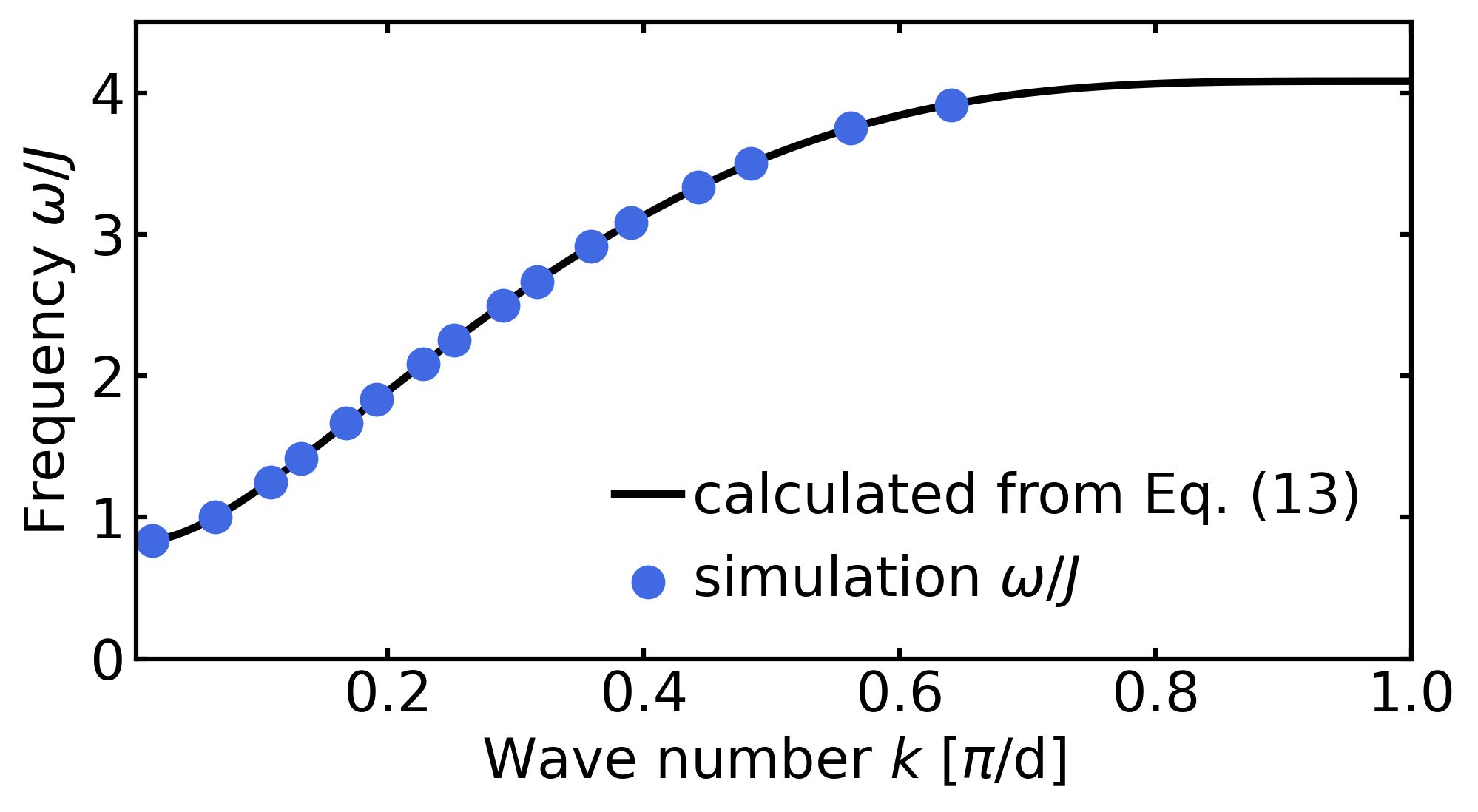}}
		\caption{(color online) Dispersion relation $\omega(k)$ in units of $J$ of an AFM spin wave where the dots represent the fitted data from the simulations (fitted $k$ to the corresponding frequency $\omega$) and the solid line shows the frequency calculated using Eq.\ \eqref{eq:disp_rel}.
			\label{fig:disp_rel}}
	\end{figure}

\begin{figure}[t!]
		\subfloat{\includegraphics[width=0.48\textwidth]{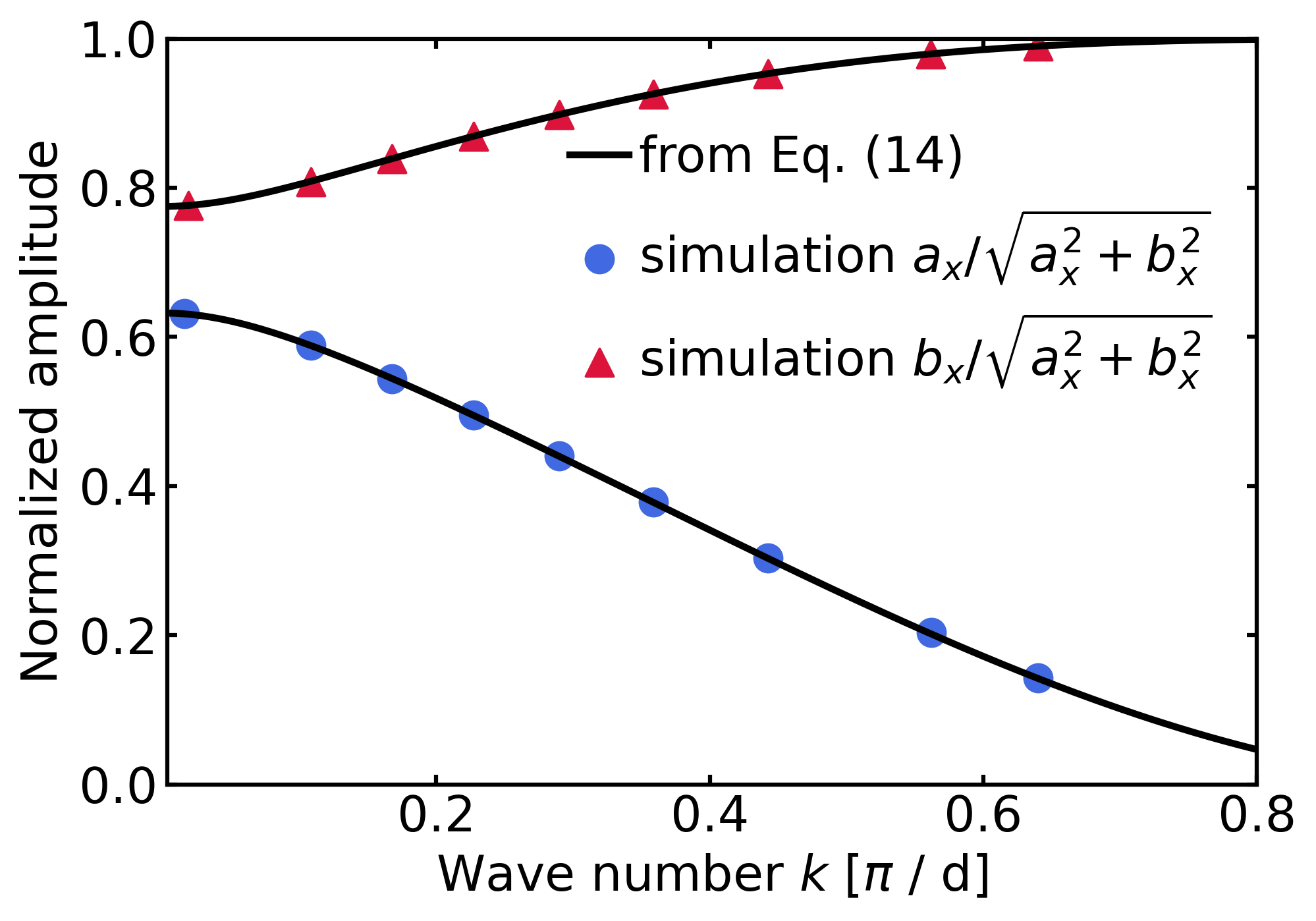}}
		\caption{(color online) Normalized amplitudes of a circularly polarized spin wave on sublattice A (blue) and sublattice B (red). The symbols mark the fitted data while the solid lines are calculated from the eigenvector $\bm{\nu}$ as in Eq.\ \eqref{eq:eig_vec_exp}. The parameters are $D/J=0$, $K/J=0.4$, and $A_0 = 0.05 M_S$.
		\label{fig:eig_vecs_circ}}
	\end{figure}

\begin{figure}[t!]
		\subfloat{\includegraphics[width=0.48\textwidth]{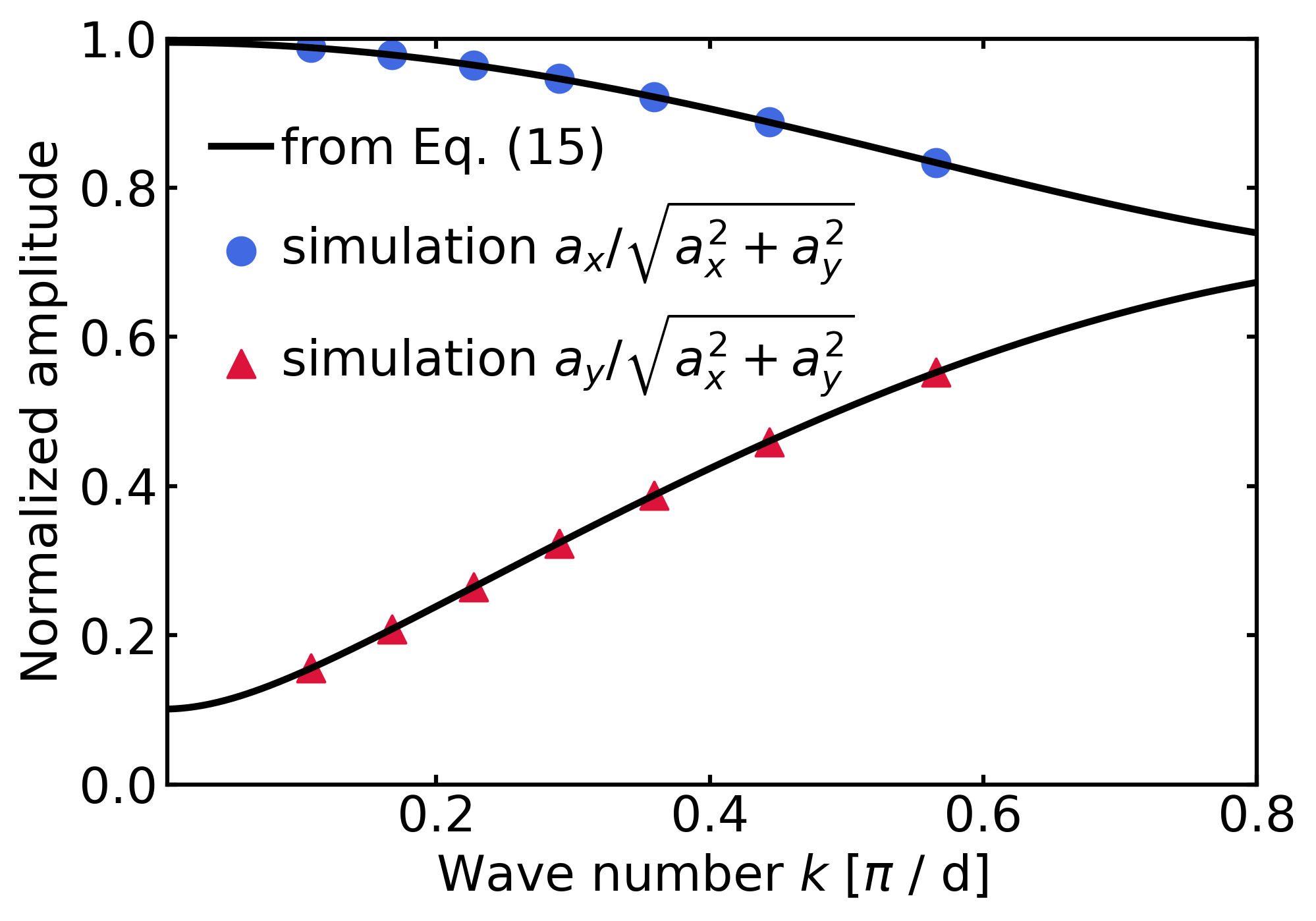}}
		\caption{(color online) Normalized amplitudes of a linearly polarized spin wave for the $x$-components (blue) and the $y$-components (red) of the spin wave. The symbols mark the fitted data while the solid lines are calculated from the eigenvector components as in Eq.\ \eqref{eq:eig_vec_exp}. The parameters are $D/J=0$, $K/J=0.4$, and $A_0 = 0.05 M_S$. 
		\label{fig:eig_vecs_lin}}
	\end{figure}

	In order to verify the analytical results of the preceding section, we next compare them to simulation data, based on Eq.\ \eqref{eq:Hamiltonian_general}. To this end, we prepare the lattice in the classical AFM ground state and drive the magnetic moments on the left edge of the lattice into oscillation by external forcing. Edge spin manipulation is a common tool when working with spin lattices \cite{SieglPRB22,SchaefferSciRep20}. To create circularly polarized spin waves, we oscillate the leftmost spins, e.g.,  $\bm{a} = (A_0 \cos(\omega t), A_0 \sin(\omega t), \sqrt{1-A_0^2})^T$, with a chosen frequency $\omega$ and an amplitude $A_0/M_S = 0.05$. In order to create linearly polarized waves, the spins are oscillated similarly, however with only one component oscillating, e.g., $\bm{a} = (A_0 \cos(\omega t), 0, \sqrt{1-A_0^2 \cos^2(\omega t)})^T$. After a few oscillations, the spin wave travels long enough through the lattice in order to obtain a sufficient amount of data points to fit the components of the spins to sine functions. 
	\begin{figure}[t!]
		\subfloat{\includegraphics[width=0.48\textwidth]{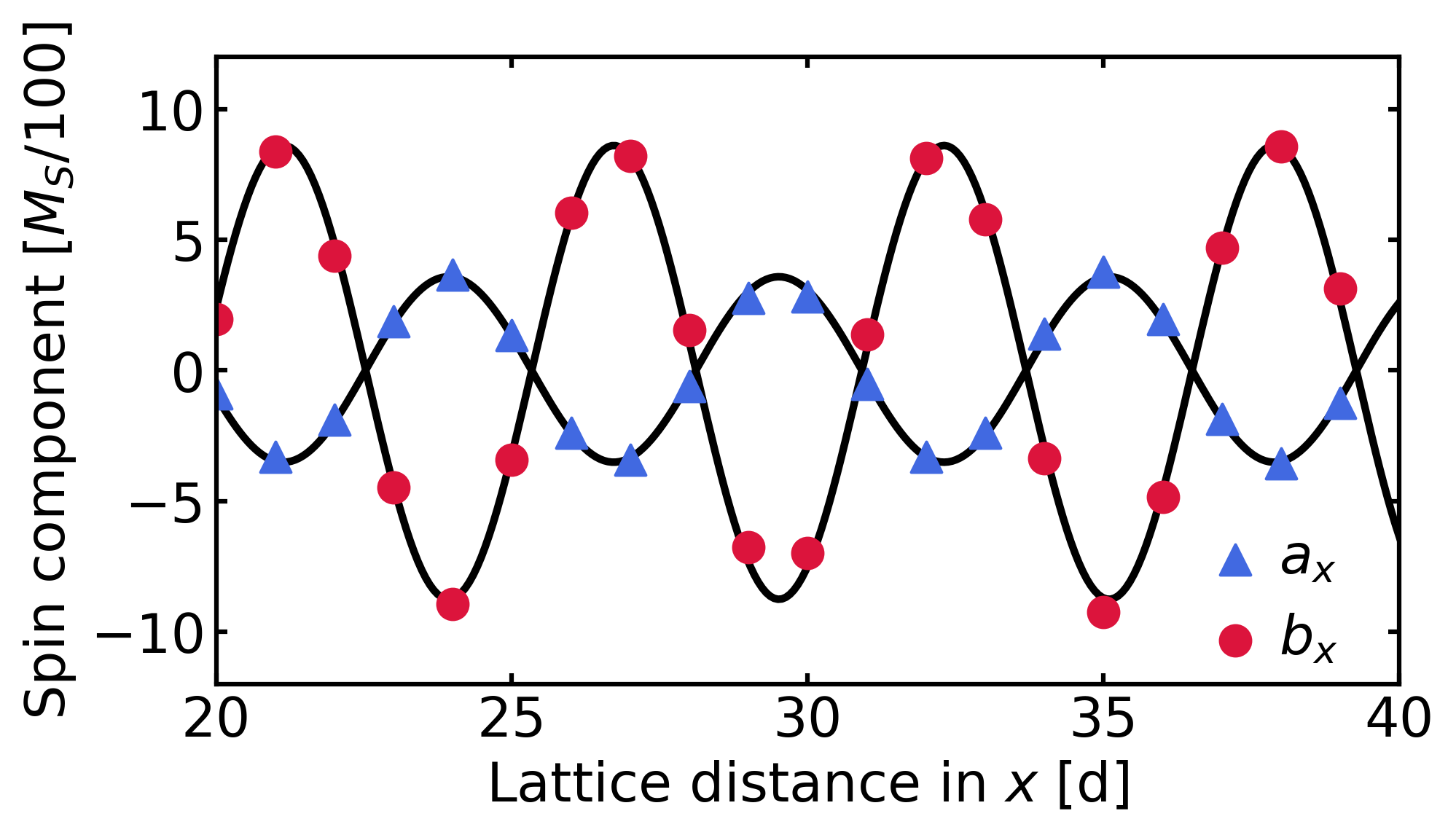}}
		\caption{(color online) Snapshot of the components $a_x$ and $b_x$ of the spins along the cross section in the $x$-direction of the lattice as a function of the distance to the left edge are plotted as colored symbols. The solid lines are the corresponding fitted sine functions. The parameters are  $K/J=0.04$ and $A_0/M_S = 0.05$. The wave is induced by rotating the spins at the edge $x=0$ of the lattice in time. 
			\label{fig:fit_wave}}
	\end{figure}
	In Fig.\ \ref{fig:fit_wave}, the $x$-component of the magnetic vectors $\bm{a}$ and $\bm{b}$ along the lattice in $x$-direction are plotted together with the corresponding fitted sine functions. Because these sine functions are along the lattice, the fit gives the lattice vector $k_x$ as well as the amplitude for each spin component in the simulation. Note that, in order to get a controlled spin wave, we construct the simulations so that $k_y=0$, which in turn means that the considered wave number is only in $x$ direction $k := k_x$. Hence, we have tuples of data points $(\omega, k, a_x, b_x, a_y, b_y)$ which we can compare to the analytical results. All simulations were made with the system parameter $K/J=0.04$. The dispersion relation $\omega(k)$ is shown in Fig.\ \ref{fig:disp_rel}. The simulated data fit well to the analytical results of Eq.\ \eqref{eq:disp_rel}. It also reproduces the well-known linear slope for medium large $k$ or if $K \rightarrow 0$, as well as the gap which is due to the easy-axis term $K$ in the $z$-direction \cite{fazekas1999lecture}.
	Additionally, we calculate the normalized eigenvectors (amplitudes) for both, linearly and circularly polarized waves, from the data by normalizing the fitted amplitudes and interpreting them as a normalized eigenvector
	\begin{equation}
		\bm{\psi}_{\pm} = \begin{pmatrix}
			a_x \\
			b_x \\
			a_y \\
			b_y
		\end{pmatrix} \big/ \sqrt{a_x^2+b_x^2+a_y^2+b_y^2} \, .
	\end{equation} 
	These data and the comparison to the analytical results are shown in Figs.\ \ref{fig:eig_vecs_circ} and  \ref{fig:eig_vecs_lin}. The simulations confirm the analytical calculations. They also perfectly reproduce  the sublattice symmetry of the circularly polarized spin waves and the $x$-$y$-symmetry of the linearly polarized spin waves, respectively. In addition, we have also investigated different values of the DMI $D$ in the range of $D/J \in [0; 0.15]$ and found no significant impact on the spin waves, just like the theory predicted (data not shown). 

	\section{Skyrmion - Spin Wave Interaction}
	\label{sec:sk-sw}
	We next consider the interaction of spin waves with AFM Skyrmions. To simulate the Skyrmion-spin wave interaction, we prepare the AFM lattice with a single, isolated and stationary Skyrmion located close to the left edge of the lattice. We consider both, Skyrmions of the Bloch and of the N{\'e}el type, and find no difference in the case of circularly polarized spin waves. For all simulations reported below, we have used the system parameters $D/J=0.15$ and $K/J=0.04$. Additionally, we also performed simulations with other system parameters resulting in moving Skyrmions with different sizes. These results can be found in the appendix. Then, we inject a monochromatic spin wave into the lattice by driving the spins of the leftmost edge of the lattice into oscillation (see Sec.\ \ref{sec:num_sw}). The spin wave travels in positive $x$-direction, from left to right, over the lattice and hits the Skyrmion where it scatters. This scattering induces a net Skyrmion motion, depending on the spin wave attributes. 
	\begin{figure}[t!]
		\subfloat{\includegraphics[width=0.48\textwidth]{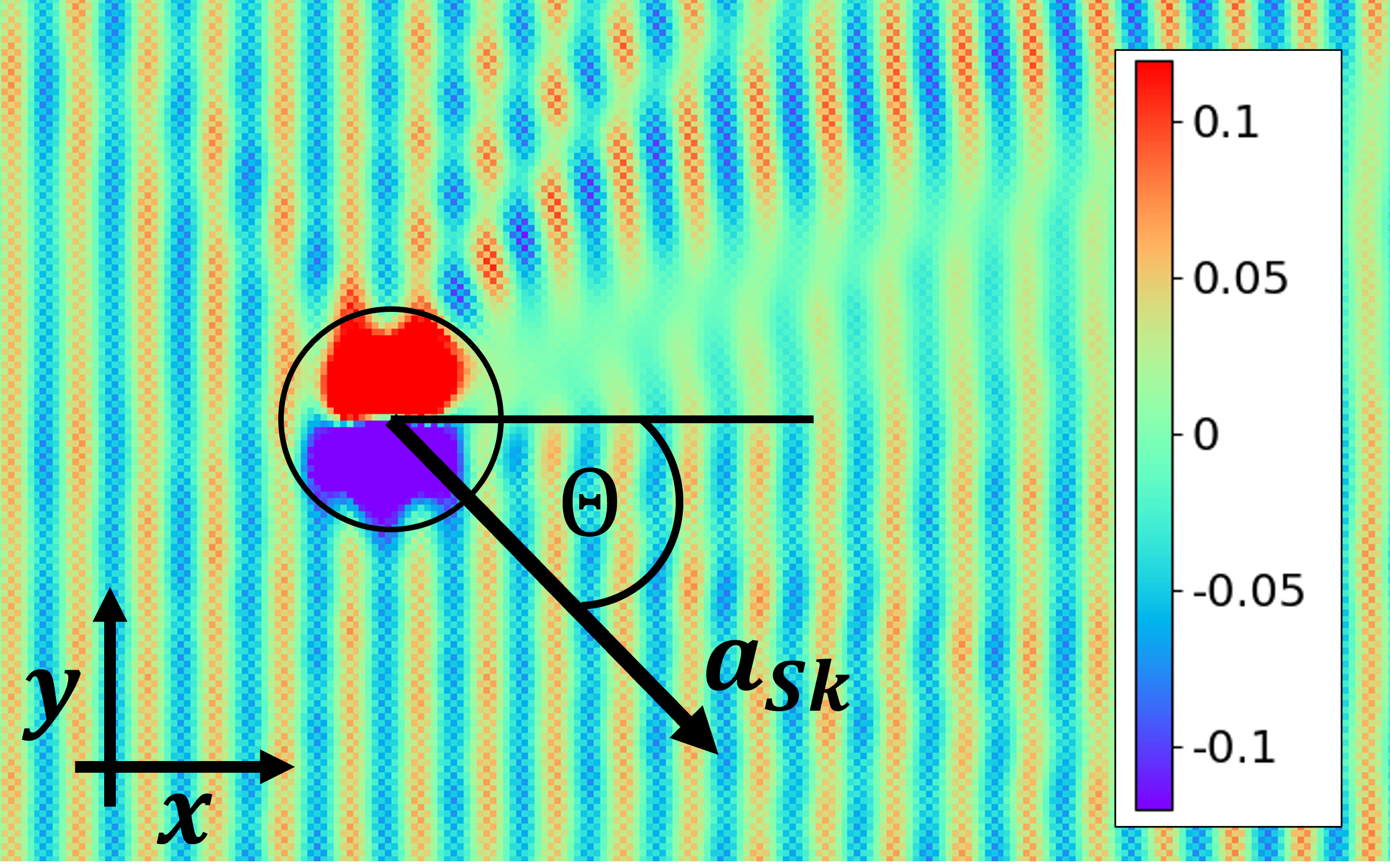}}
		\caption{(color online) Illustration of a circularly polarized AFM spin wave which scatters at an AFM Skyrmion. Shown here is the $x$ component $M_x$ of each lattice spin, cut off at $\pm 0.12 M_S$ for the benefit of better visibility. For this illustration we prepared the lattice by multiplying each spin of the sublattice B by $-1$, so that the AFM Skyrmion is visible as a coherent patch,   highlighted by the black circle. Additionally, this preparation  directly reveals the different spin wave amplitudes of the two sublattices. The spin wave propagates in the positive $x$ direction, from left to right, and scatters at the Skyrmion. This scattering induces a Skyrmion acceleration $a_{\text{Sk}}$, depicted here by an arrow. Circularly polarized spin waves accelerate the Skyrmion with a Skyrmion Hall angle  $\Theta$. The parameters are $D/J=0.15$, $K/J=0.04$, $\omega/J= 1.8$ and the driving amplitude $A_0 = 0.05 M_S$.
			\label{fig:2d_sk_sw}}
	\end{figure}
	As a most important difference to ferromagnetic spin wave driven Skyrmions, which propagate against the spin wave propagation direction \cite{schuette,sw_iwasaki,schroeter_garst}, the AFM spin waves drive the AFM Skyrmion into their propagation direction. They induce a force which accelerates the Skyrmion. This acceleration is common for AFM structures influenced by a force \cite{tveten2013staggered}. In passing, we note that we have also simulated ferromagnetic spin wave - Skyrmion scattering and found that this results in much slower Skyrmion motion than in the AFM case (see appendix for more details). In Fig.\ \ref{fig:2d_sk_sw}, the scattering of an AFM spin wave at the Skyrmion is illustrated for a specific example. The presentation of the data is such that one can compare it with studies of ferromagnetic spin wave Skyrmion scattering \cite{schuette}. Movies of the corresponding scattering of circularly and linearly polarized spin waves with Skyrmions and subsequent Skyrmion motion are included in the Supplemental Material \cite{supp1}. Of central interest is the time-dependent position of the Skyrmion center during the simulation. 
	
	For a quantitative analysis of the motion of the Skyrmion, we determine its position by multiplying each sublattice B spin by $-1$ to obtain a ferromagnetic Skyrmion representation, then calculate the center of mass of the topological charge density according to Ref.\ \cite{stier_strobel}. Furthermore, we compare effects of circularly and linearly polarized spin waves in the following.
	
	\subsection{Circularly Polarized Spin Waves}
\begin{figure}[t!]
		\subfloat{\includegraphics[width=0.48\textwidth]{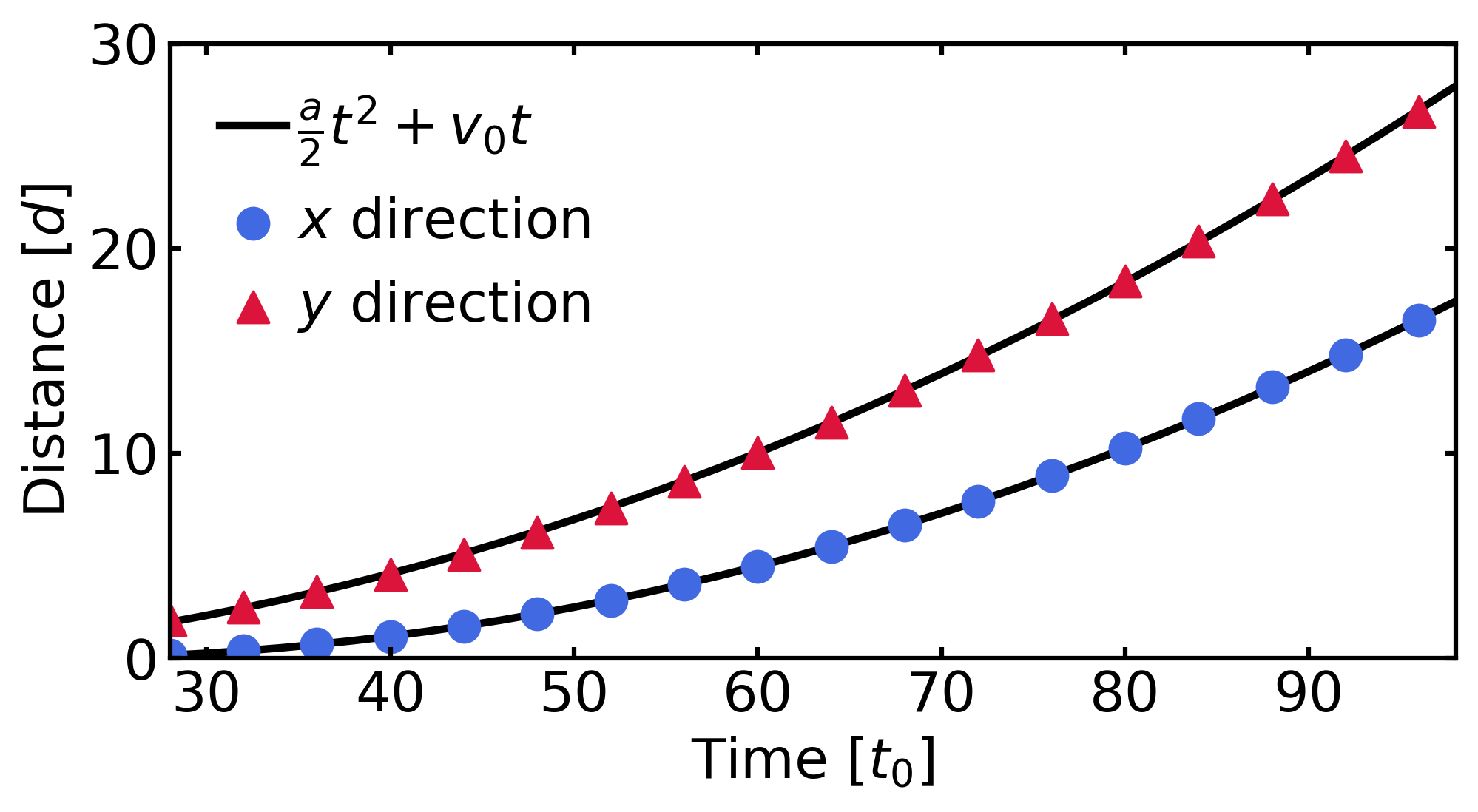}}
		\caption{(color online) Distance traveled by the AFM Skyrmion (in units of the lattice constant $d$) over time (in units of $t_0 = 1/J$) while it is under the impact of a circularly polarized spin wave. The symbols represent the simulated data while the solid lines are fitted square functions to these data points. Because circularly polarized spin waves induce a Skyrmion Hall effect, the plot shows $x$- and $y$-directions. The data were simulated with the system parameters $D/J = 0.15$ and $K/J = 0.04$, and spin wave parameters $\omega/J=1.8$ and $A_0 = 0.02 M_S$.
			\label{fig:sk_motion_over_time_circ}}
	\end{figure}

	We find that when a circularly polarized AFM spin wave scatters at a Skyrmion of either type, Bloch or N{\'e}el, it moves the Skyrmion not only in the same direction of the spin wave propagation ($x$-direction), but also induces a perpendicular motion ($y$-direction). In Fig.\ \ref{fig:sk_motion_over_time_circ}, the distance is shown, in units of lattice sites $d$, traveled by the Skyrmion in $x$- and in $y$-direction, respectively, over time. 
	The symbols represent the position of the Skyrmion center at the corresponding times. The numerical data suggest an accelerated Skyrmion motion which is also consistent with the finding that AFM Skyrmions behave like massive particles in a mean-field picture \cite{velkov}. Therefore, we fit a square function $f(t) = \frac{a_{\text{Sk}}}{2} t^2 + v_0 t + s_0$, where the time $t$ is in units of $t_0 = 1/J$, to the data and extract the Skyrmion acceleration $a_{\text{Sk}}$ as a quantifier for the force acting on the Skyrmion since $F = m_{\text{eff}} a_{\text{Sk}}$. 
	
\begin{figure}[t!]
		\subfloat{\includegraphics[width=0.48\textwidth]{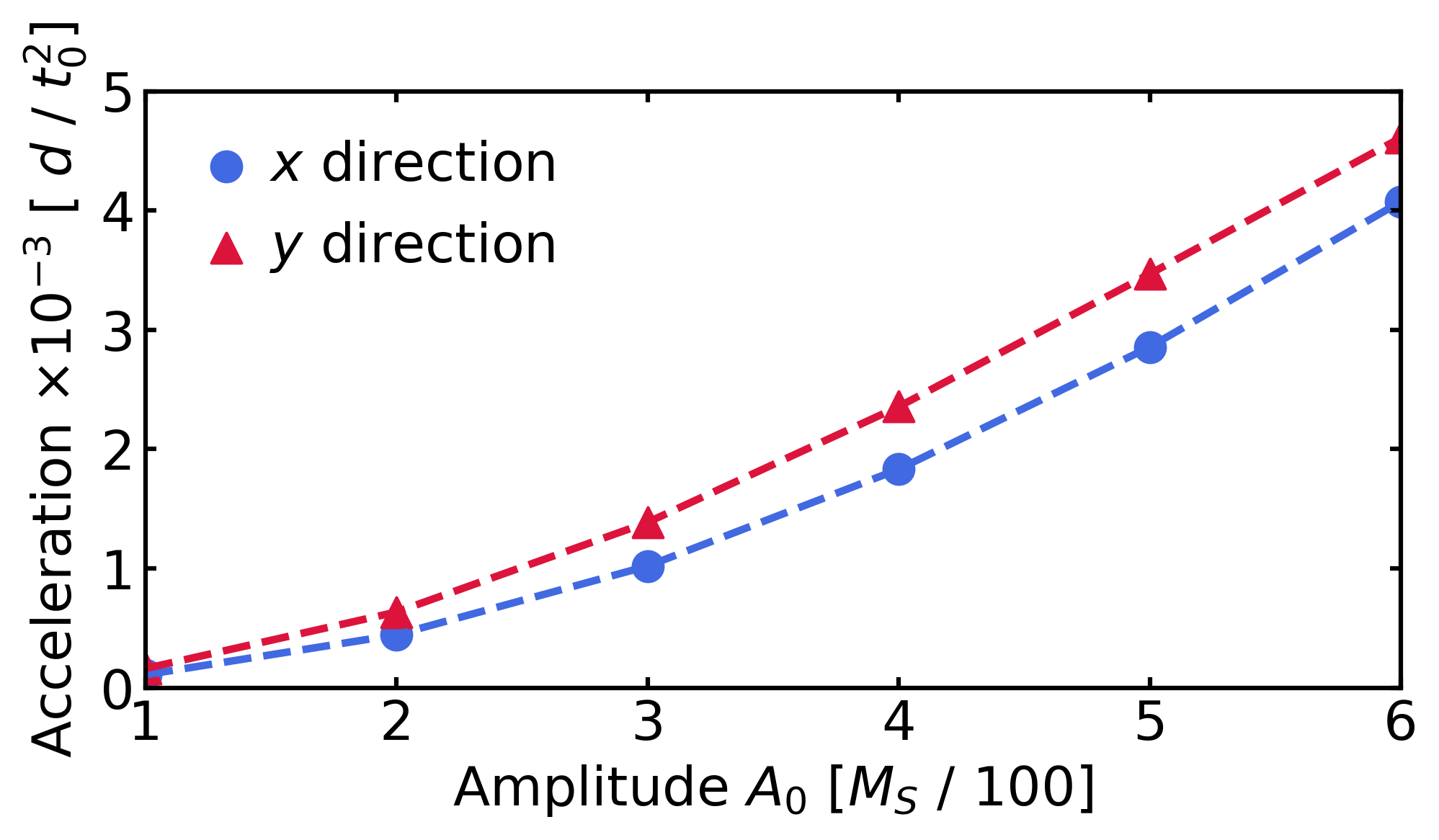}}
		\caption{(color online) Longitudinal ($x$, blue circles) and transversal ($y$, red triangles) acceleration of an AFM Skyrmion due to circularly polarized spin waves with different driving amplitudes $A_0$. The data were extracted from simulated trajectories of the Skyrmion center for  the parameters $\omega/J=1.8$, $D/J = 0.15$, and $K/J = 0.04$. 
			\label{fig:sk_acc_over_amp_circ}}
	\end{figure}

\begin{figure}[t!]
		\subfloat{\includegraphics[width=0.48\textwidth]{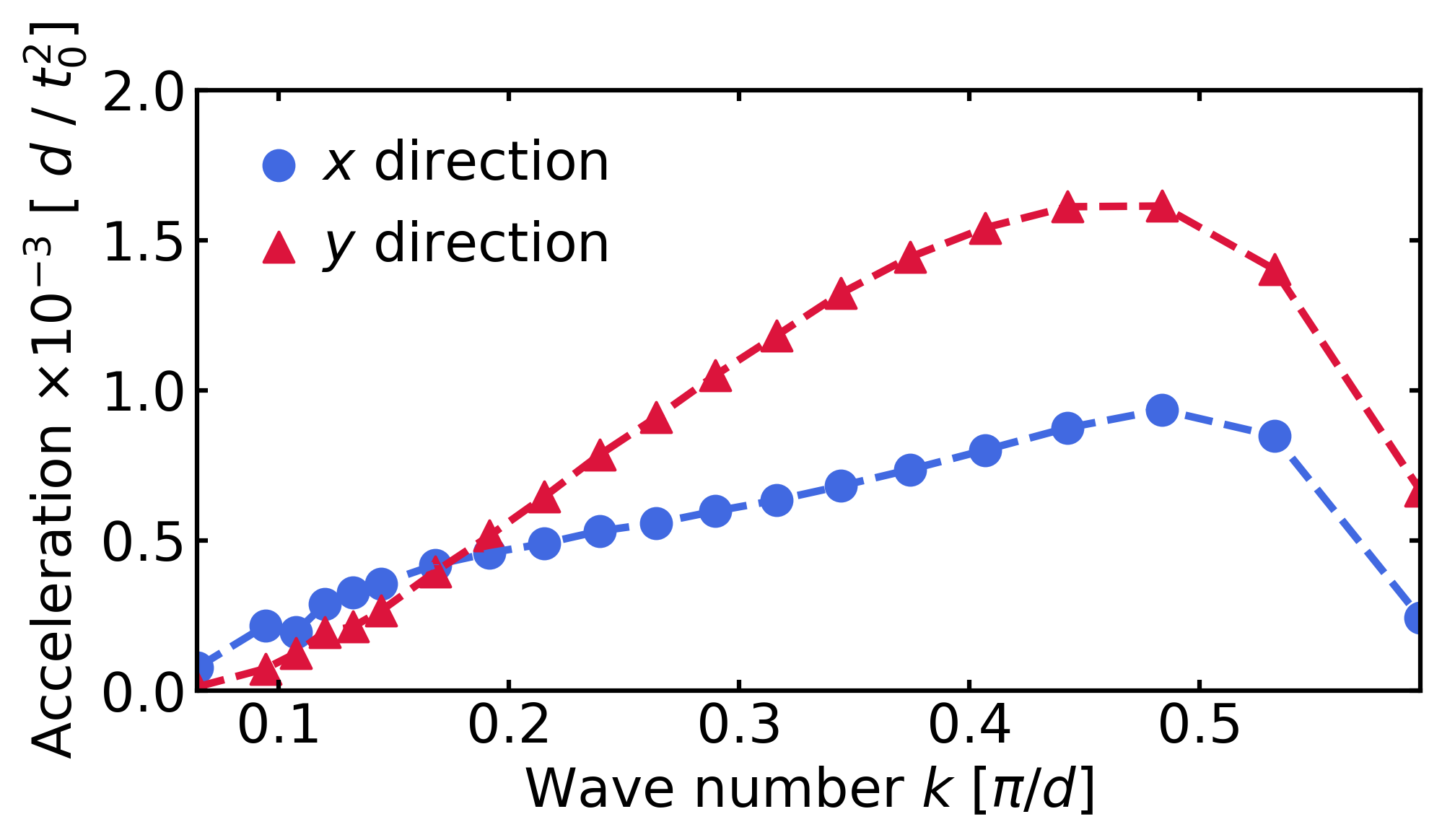}}
		\caption{(color online) Longitudinal ($x$, blue circles) and transversal ($y$, red triangles) acceleration of the Skyrmion due to circularly polarized spin waves with different wave numbers $k$. The parameters are $D/J = 0.15$ and $K/J=0.04$. The spin wave is generated with a driving amplitude $A_0 = 0.02 M_S$.
			\label{fig:sk_acc_over_k_circ}}
	\end{figure}

	The magnitude of the Skyrmion acceleration induced by the spin wave strongly depends on the spin wave amplitude. The latter is a control parameter of the external drive. To analyze this dependence, we have simulated the spin wave - Skyrmion scattering with a variety of spin waves of same frequency $\omega/J = 1.8$, but different oscillation amplitudes $A_0$. For each simulation the traveled distance of the Skyrmion center over time was fitted to a square function. The resulting accelerations in the $x$- and $y$-direction, respectively, are plotted against the spin wave driving amplitude in Fig.\ \ref{fig:sk_acc_over_amp_circ}. As expected, spin waves with larger amplitudes accelerate the Skyrmion stronger. At larger Skyrmion velocities or at stronger driving it was no longer possible to fit the track of the Skyrmion center over time to a square function and thus to a constant acceleration. Other effects start to emerge. For instance, the Doppler effect could arise so that the moving Skyrmion experiences a spin wave with an effectively larger wave length. This lowers the acceleration, see Fig.\ \ref{fig:sk_acc_over_amp_circ}, and thus reduces the velocity as compared to a uniform acceleration.	Such effects leading to a non-constant Skyrmion acceleration are not shown here since analyzing this more complicated motion is outside the scope of this work. We consider only small spin wave amplitudes of at most two percent of the spin length, so that the resulting Skyrmion motion is slow enough to be fitted against a constant acceleration. In addition, the assumptions necessary for the linearized spin wave theory of Sec.\ \ref{sec:spinwave} to apply are fulfilled by such small spin wave amplitudes. 
	Nevertheless, acceleration of Skyrmions not only depends on the spin wave driving amplitude, but also on the wave number $k$ of the spin wave. This is shown in Fig.\ \ref{fig:sk_acc_over_k_circ} where we depict the fitted Skyrmion acceleration in the longitudinal ($x$) and transversal ($y$) direction, respectively, versus $k$.
	Both the accelerations first grow with growing wave numbers until, however, an upper bound is reached above which  the Skyrmion acceleration drops rapidly. Additionally, the dependence of the Skyrmion acceleration on the wave number is different for the $x$ and $y$ direction. Hence, there exists a SHE for spin wave driven Skyrmions.

\begin{figure}[t!]
		\subfloat{\includegraphics[width=0.48\textwidth]{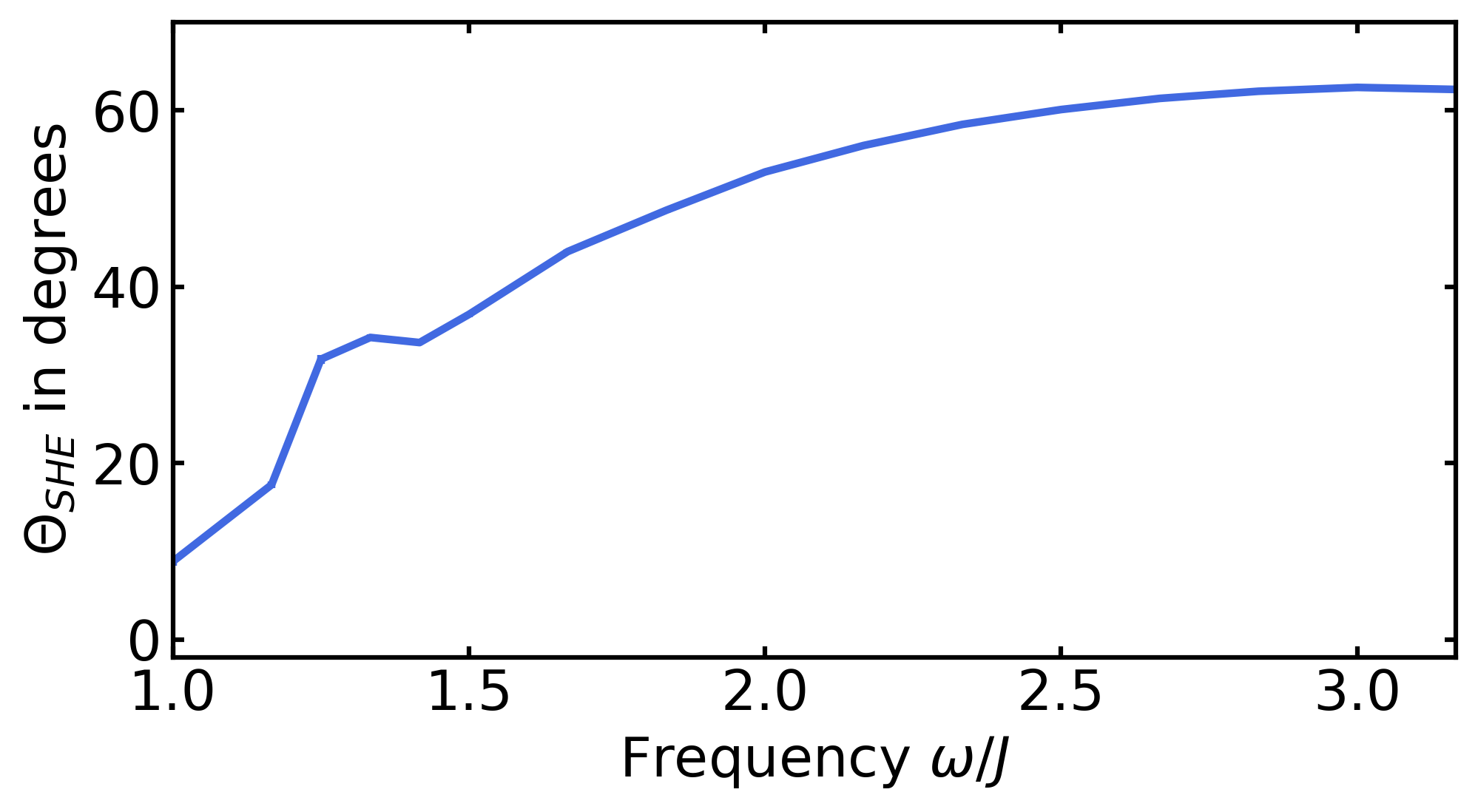}}
		\caption{(color online) Skyrmion Hall angle $\Theta_{\text{SHE}}$ versus spin wave frequency $\omega$, shown in degrees for better comparison with Ref.\ \cite{daniels_sk}. We obtain a similar behavior even though we study a bipartite AFM on a lattice instead of a synthetic AFM in the continuum approximation. We have used the parameters $D/J = 0.15$ and $K/J=0.04$.
			\label{fig:she}}
	\end{figure}

	An interesting property is that the spin wave induced SHE angle $\Theta_{\text{SHE}}$ is, similar to its ferromagnetic pendant \cite{schuette}, dependent on the wave number $k$ of the spin wave and thus on the driving frequency $\omega$. In Fig.\ \ref{fig:she}, the Hall angle is shown for different values of the driving frequency $\omega$. We plot the Skrymion Hall angle in degree against  $\omega$, for comparison with Ref.\ \cite{daniels_sk}. There, a similar spin wave induced Hall angle is determined for different driving frequencies, but for a synthetic antiferromagnet in the continuum approximation, while we study a bipartite AFM on a lattice. Nevertheless, the frequency dependencies of $\Theta_{\text{SHE}}(\omega)$ obtained in this work appear as qualitatively similar to that of Ref.\ \cite{daniels_sk}. In addition, we find a non-monotonous dependence of the Skyrmion Hall angle on the frequency  in Fig.\ \ref{fig:she}  in the low-frequency regime. This effect is stronger for larger Skyrmions, which we show in the Appendix. Even though the non-monotonous ``hump'' in the SHE is visible, one should keep in mind that the absolute acceleration of the Skyrmion in this regime is rather small (cf. Fig.\ \ref{fig:sk_acc_over_k_circ}).

	Regarding the force in the longitudinal direction of the spin wave propagation, we find the same magnitudes for left- and right-handed circularly polarized spin waves $F_{\parallel}(\omega) = F_{\parallel}(-\omega)$. However, in transversal direction (perpendicular to the direction of the spin wave propagation) the force reverses with handedness, $F_{\perp}(\omega) =-F_{\perp}(-\omega)$. Hence, the sign of the Skyrmion Hall angle depends on whether we consider left- or right handed spin waves, reflecting the sublattice symmetry (cf.\ Sec.\ \ref{sec:spinwave}) of AFM spin waves. 
	
	We conclude that the spin wave induced SHE roots in the different amplitudes the spin wave exhibits on each sublattice. Regarding each sublattice separately, they form two ferromagnetic Skyrmions with opposite topological charge $Q_A = -Q_B$. It is known that the sign of the SHE in ferromagnetic Skyrmion motion (e.g., the current-induced SHE) depends on the sign of the topological charge \cite{StierPRL17,jiang2017direct}. That means the Skyrmion experiences a different SHE on each of the both sublattices. This matches with the observation that an interchanging of the sublattices $\{A,B\} \rightarrow \{B,A\}$ results in a change of the Skyrmion Hall angle $\Theta_{\text{SHE}} \rightarrow -\Theta_{\text{SHE}}$.
	
	\subsection{Linearly Polarized Spin Waves}
\begin{figure}[t!]
		\subfloat{\includegraphics[width=0.48\textwidth]{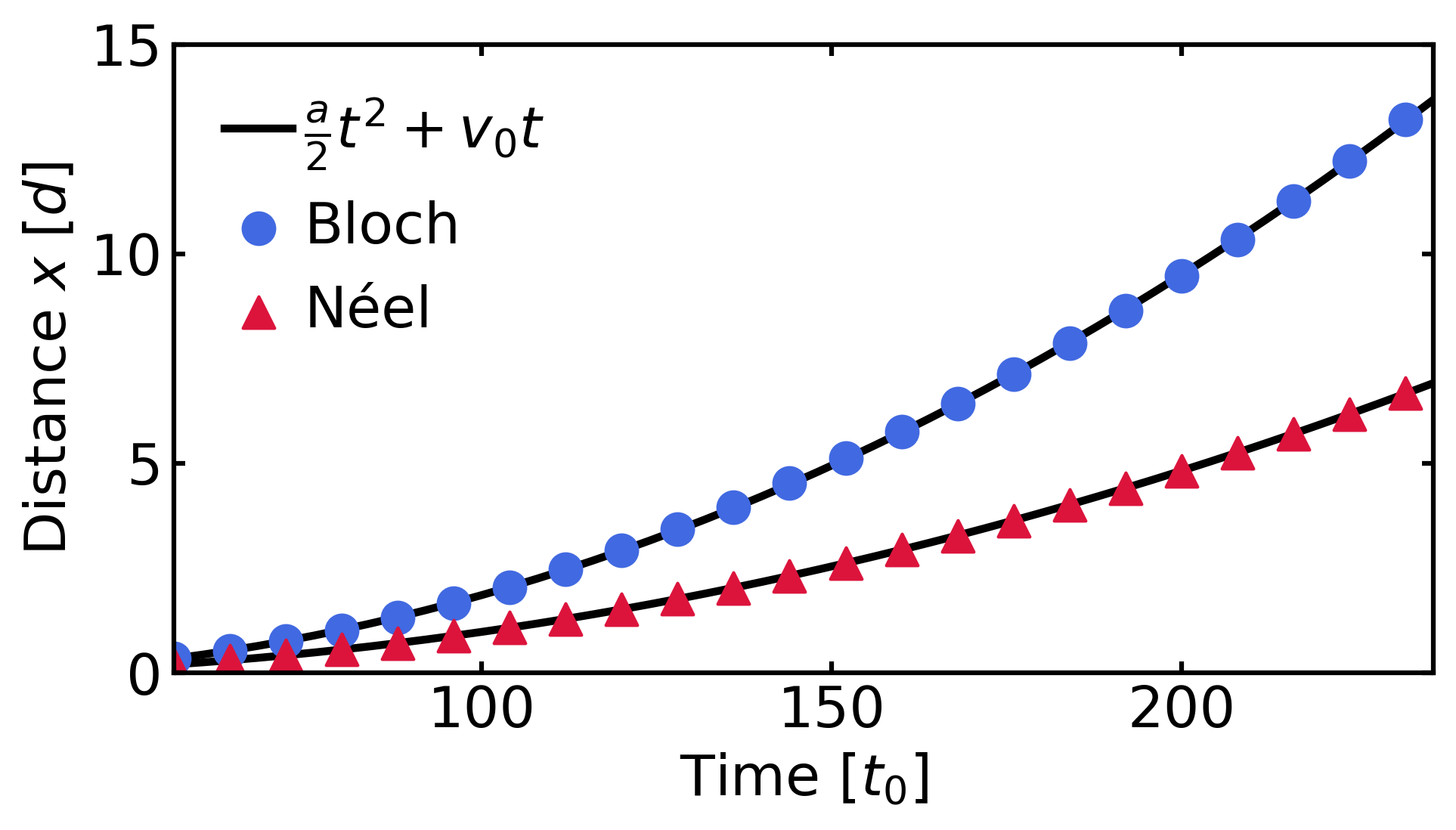}}
		\caption{(color online) Distance traveled by an AFM Bloch (blue circles) and N{\'e}el (red triangles) Skyrmion over time while being under the impact of an $x$ linearly polarized spin wave. The symbols represent the results of the simulations while the solid lines are fitted square functions to these data points. Because linearly polarized spin waves accelerate Skyrmions exclusively in longitudinal  direction, the plot only shows the traveled distance in the $x$-direction. The parameters are $D/J = 0.15$, $K/J = 0.04$, $\omega/J= 1.8$ and $A_0 = 0.05 M_S$. 
			\label{fig:sk_motion_over_time_lin}}
	\end{figure}

	For the scattering of linearly polarized spin waves at Skyrmions, we find the interesting result that they only induce a Skyrmion motion in their propagation direction with no SHE. This observation supports the finding that the SHE comes from different sublattice amplitudes, because in linearly polarized spin waves, both amplitudes are equal on the sublattices. Hence, we only may focus at Skyrmion motions in the $x$-direction. As another peculiarity, we observe that linearly polarized spin waves impact N{\'e}el and Bloch Skyrmions differently, contrary to circularly polarized waves. Both N{\'e}el and Bloch type Skyrmions are topologically protected, while Bloch Skyrmions are energetically favored by a bulk DMI and N{\'e}el Skyrmions are favored by an interfacial DMI. Theoretically it is possible to smoothly transform one type into the other without changing the topology by $(M_x,M_y,M_z) \rightarrow (-M_y, M_x, M_z)$. In Fig.\ \ref{fig:sk_motion_over_time_lin}, as one exemplary simulation, we compare the distance of Bloch- versus N{\'e}el-Skyrmions, driven by an $x$-linearly polarized spin wave, as they travel in $x$-direction. 
	The symbols mark the data obtained from the numerical simulations and the solid lines are the corresponding fits to a square function. It shows that both types of Skyrmions are accelerated, but with different magnitude. 
	
\begin{figure}[t!]
		\subfloat{\includegraphics[width=0.48\textwidth]{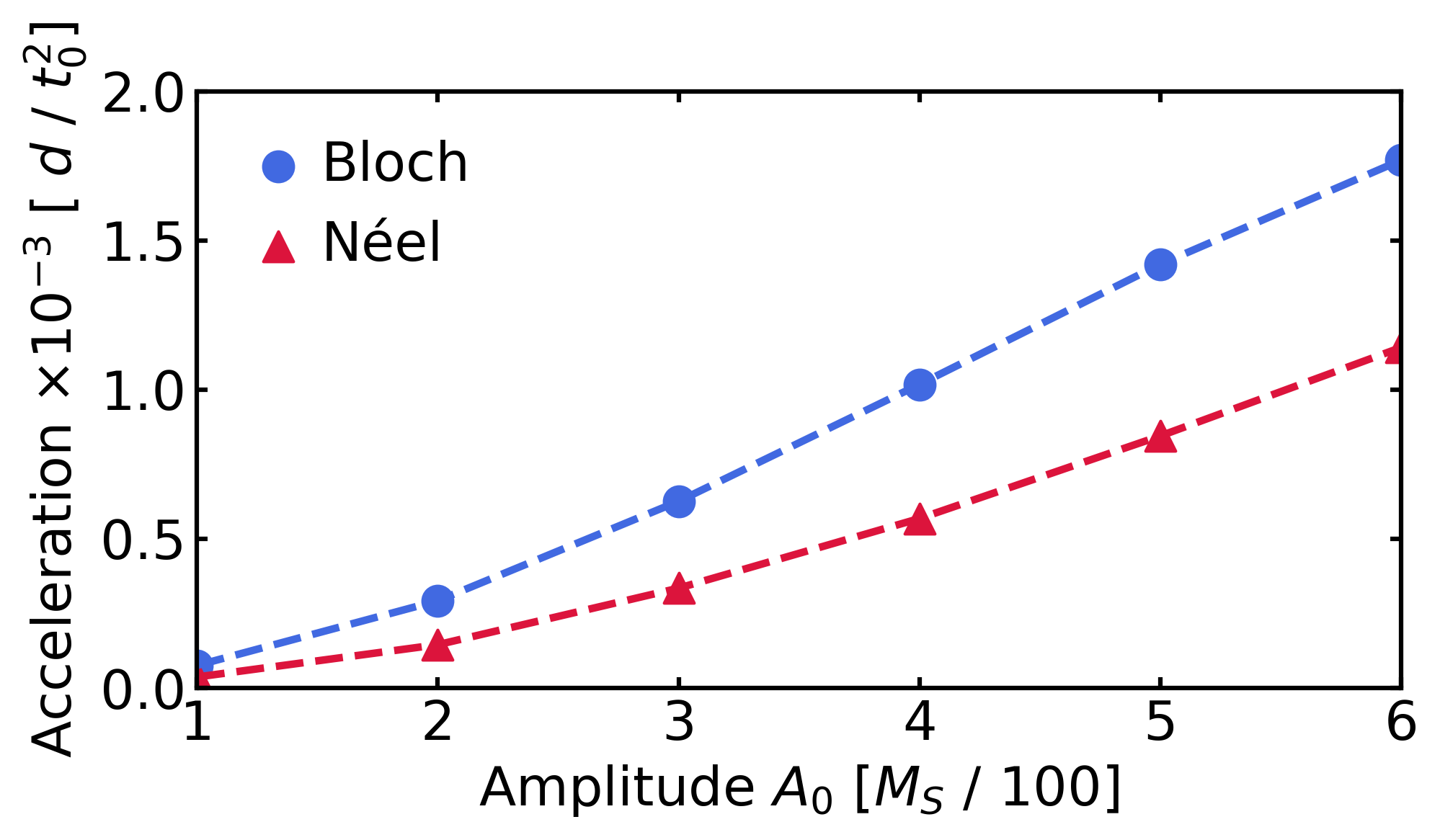}}
		\caption{(color online) Longitudinal acceleration of an AFM Bloch (blue circles) and N{\'e}el (red triangles) Skyrmion extracted from fits of the traveled distance of the Skyrmion over time (cf.\ Fig.\ \ref{fig:sk_motion_over_time_lin}) for different simulations using linearly polarized spin waves with the same frequency $\omega / J = 1.8$. 
			\label{fig:sk_acc_over_amp_lin}}
	\end{figure}

\begin{figure}[t!]
		\subfloat{\includegraphics[width=0.48\textwidth]{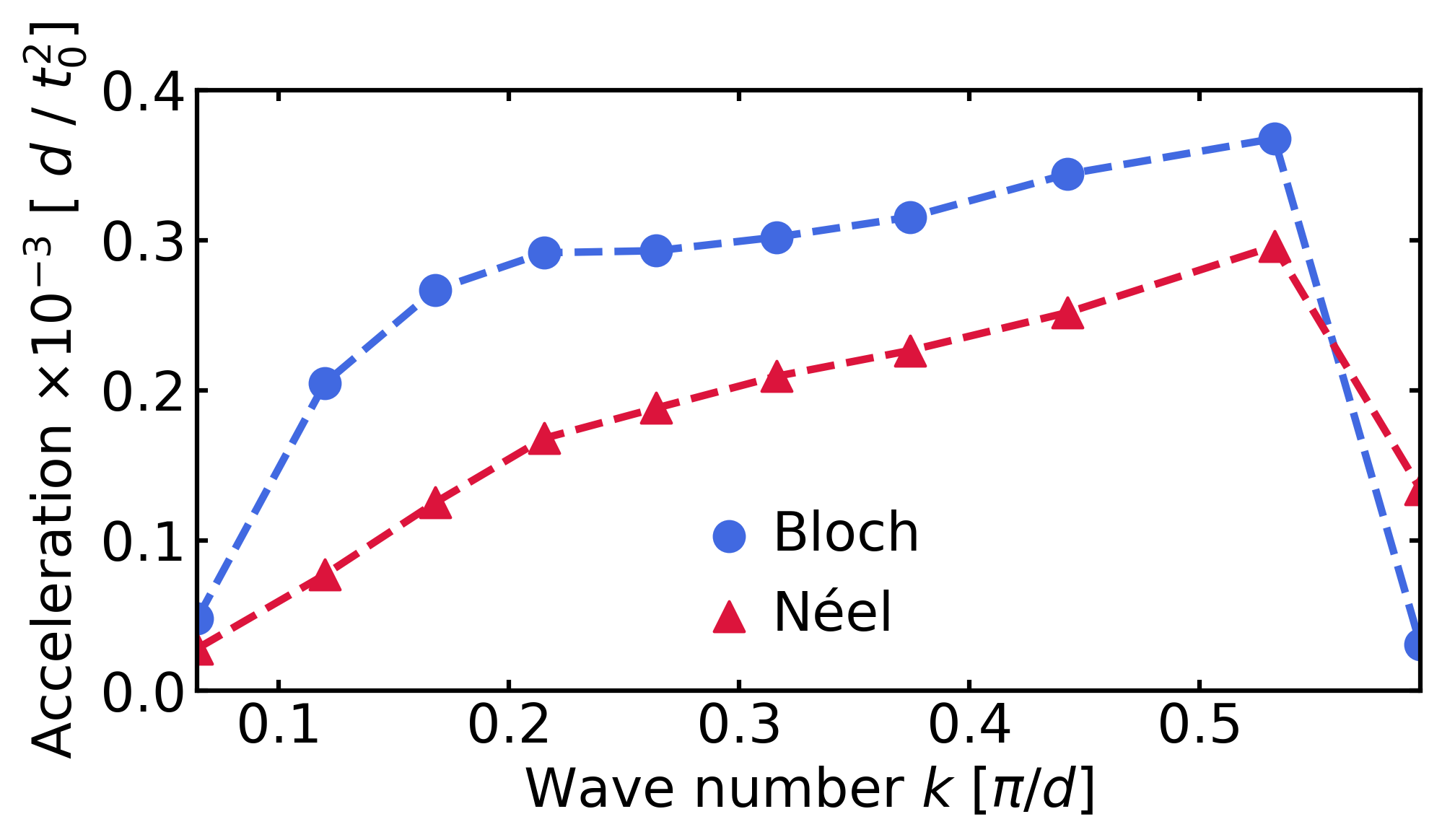}}
		\caption{(color online) Longitudinal acceleration of a Bloch Skyrmion (blue) and a N{\'e}el Skyrmion (red) due to  linearly polarized spin waves with different wave numbers $k$. The parameters are $D/J = 0.15$ and $K/J=0.04$, where the spin wave was generated with a driving amplitude $A_0 = 0.02 M_S$.
			\label{fig:sk_acc_over_k}}
	\end{figure}

	We may extract the magnitudes of the accelerations from the fits and interpret them as force acting on the Skyrmion. As for the circularly polarized spin waves, this force strongly depends on the spin wave driving amplitudes. In Fig.\ \ref{fig:sk_acc_over_amp_lin}, the Skyrmion acceleration (in longitudinal direction) is plotted against varying spin wave amplitudes for Bloch- and N{\'e}el Skyrmions. We observe that the larger the amplitude is, the larger the acceleration of each of these becomes, but also the difference increases.
	As for the circularly polarized spin waves (see the previous section), the (longitudinal) acceleration of Skyrmions by linearly polarized spin waves also depends on the wave number $k$. We plot the acceleration of an AFM Bloch (blue circles) and N{\'e}el (red triangles) Skyrmion as a function of the wave number in Fig.\ \ref{fig:sk_acc_over_k}. 
	Regardless of the wave number, Bloch Skyrmions are in general accelerated stronger by $x$ linearly polarized spin waves than N{\'e}el Skyrmions. Both Skyrmion types accelerate stronger for higher wave numbers until an upper limit, however, at which this effect drops rapidly for wave numbers $k \gtrsim \pi/(2d)$.
	
	The previously mentioned transformation between Bloch- and N{\'e}el Skyrmions is similar to the transformation from $x$- to $y$ linearly polarized spin waves. Therefore, we have also studied the impact of $x$ and $y$ linearly polarized spin waves on both Bloch- and N{\'e}el Skyrmions and found that an $x \leftrightarrow y$ polarization exchange bears the same results as a Bloch $\leftrightarrow$ N{\'e}el exchange (data not shown). Thus, the component symmetry of the linearly polarized spin wave has an immediate effect on the force it applies to the Skyrmion.
		
	\section{Damped Spin Waves and Skyrmion Motion}
	\label{sec:damping}
	Above, we have considered the dynamics in the absence of any damping. In this section we report on analytical calculations for damped AFM spin waves and provide numerical results. To this end, we linearize the LLG equation \eqref{eq:LLG}, analogously to Eq.\ \eqref{eq:lin_eom_nondamped}, and include a non-zero damping term $\alpha > 0$. First, for ferromagnetic spin waves we find that damping has only a marginal effect on the dispersion relation $\omega(\bm{k}) = (4J + 2K -2J[\cos(kx) + \cos(ky)]) / (1 + \alpha^2)$ while it damps the amplitude of spin waves exponentially over time $\psi = \psi_0 \exp(-\alpha \omega t)$, depending on frequency $\omega$.
	
\begin{figure}[t!]
		\subfloat{\includegraphics[width=0.48\textwidth]{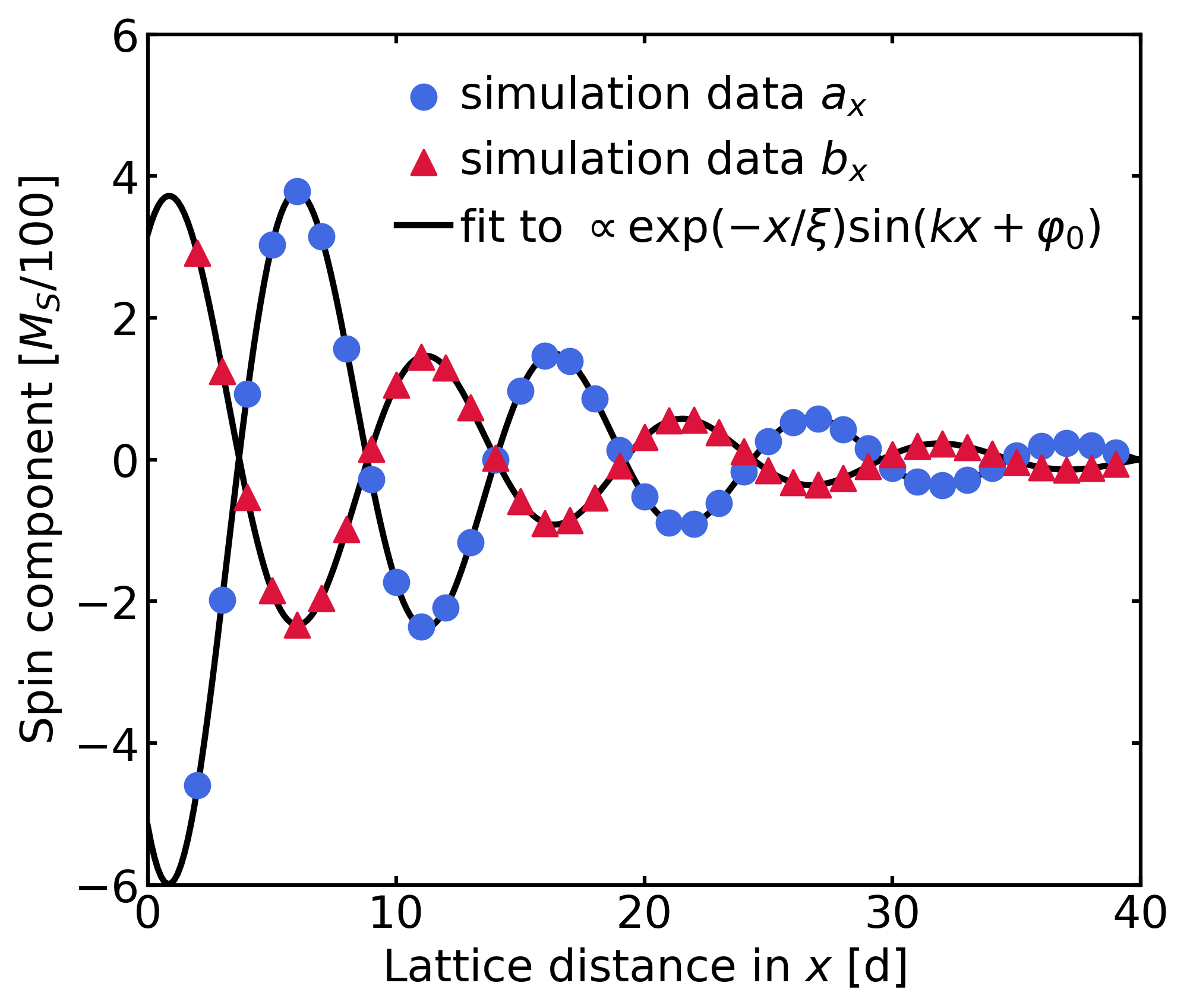}}
		\caption{(color online) Snapshot of a circularly polarized spin wave damped along the lattice in longitudinal direction. The symbols are the $x$ component of a spin wave in the sublattice A (blue circles) and sublattice B (red triangles), respectively. The solid lines are damped oscillation functions $\sim \exp(-x/\xi) \sin(kx + \varphi_0)$, fitted to this data. The parameters are $D/J=0$, $K/J=0.04$, $\omega/J = 1.8$, $A_0 = 0.05 M_S$ and $\alpha = 0.05$.
			\label{fig:example_damped_wave}}
	\end{figure}

	The AFM case is more involved and we only consider small values of  $\alpha \ll 1$. The detailed calculations are given in Appendix \ref{apendix1}. Similar to the ferromagnetic case, no significant change of the frequency appears by weak damping. However, we find a phase shift in the amplitudes between the two sublattices A and B, whose magnitude depends on the damping $\alpha$. Even though this shift is linear in $\alpha$ for small $\alpha$, it remains too small to be directly visible in our simulation results. Also similar to the ferromagnetic case, with time the amplitudes show exponential damping, albeit with one important difference: In the AFM case, the damping over time $\psi_{\text{AFM}} = \psi_0 \exp(-\alpha \rho t)$ appears to be independent of the spin wave frequency, though dependent on the system parameters, in striking contrast to the ferromagnetic case. Since we consider a lattice in real space and continuously inject a spin wave from the left edge we need to monitor the wave damping over the distance traveled, before it reaches, e.g., the Skyrmion. In Fig.\ \ref{fig:example_damped_wave}, we show this amplitude along the lattice, where the symbols mark data points obtained from the simulations for circularly polarized spin waves and the solid lines are fits to an exponentially damped oscillation $\sim \exp(-x/\xi) \sin(kx + \varphi_0)$.
	In the simulations, we drive the edge spins into oscillation and let the resulting spin wave traveling into the lattice. Therefore, we consider the spin wave amplitude over space rather than over time. To this end, we assume that the amplitude decreases over time as described above and that the spin wave travels with a constant velocity. This leads to a decreasing amplitude with increasing spatial distance. To estimate the corresponding decay length, we calculate the traveling time of the wave front $t = x/v_g$ with the group velocity derived from Eq.\ \eqref{eq:disp_rel} 
	as 
	\begin{equation}
		\label{eq:group_vel_afm}
		v_g(k) = \frac{\partial \omega}{\partial k} = \frac{4 J^2 d \sin(kd) \left[\cos(kd) + 1\right]}{\sqrt{\rho^2 - 4J^2 \left[ \cos(kd) + 1 \right]^2}} \, .
	\end{equation}
	By this, a spatial decay of the spin wave $\psi = \psi_0 \exp(-x/\xi)$ follows from $\psi_{\text{AFM}}$, with the spin wave decay length 
	\begin{equation}
		\label{eq:range_of_sw}
		\xi(k) = \frac{v_g(k)}{ \alpha \rho } \, .
	\end{equation}
	Since the damping in time is independent of the spin wave frequency (and thus of the wave number), the decay length is the group velocity $v_g$ scaled by the inverse energy $1/\rho$ and by the inverse damping $1/\alpha$. 
	
\begin{figure}[t!]
		\subfloat{\includegraphics[width=0.48\textwidth]{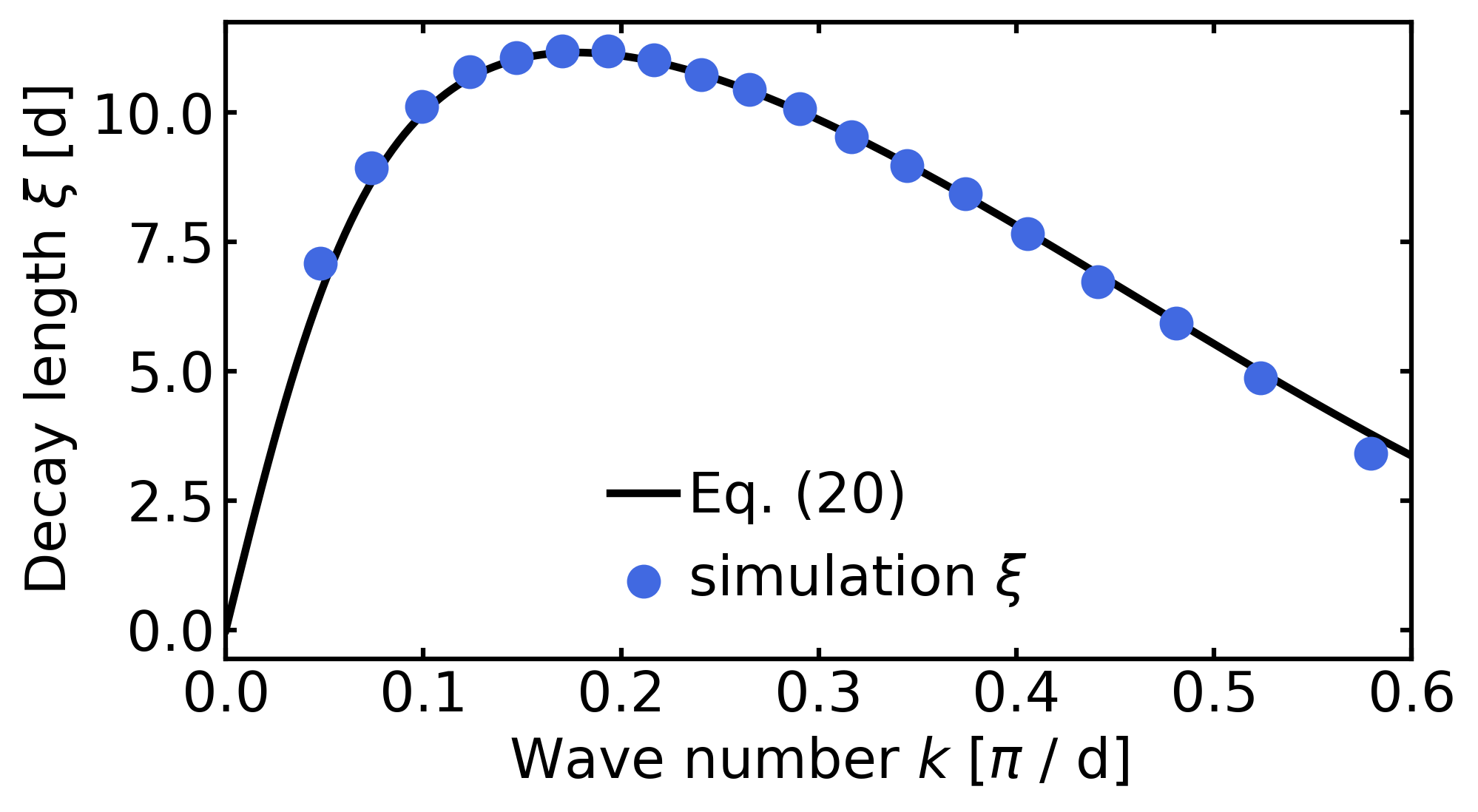}}
		\caption{(color online) Decay length $\xi$ of a damped AFM spin wave plotted against the wave number $k$. The circles represent the data points from fitting the simulation results to $\psi = \psi_0 \exp(-x/\xi)$ and the solid line is calculated using Eq.\ \eqref{eq:range_of_sw}. We have  used the parameters $D/J=0$, $K/J=0.04$ and $\alpha = 0.05$.                                
			\label{fig:decay_length}}
	\end{figure}

	To compare this result with numerical simulations, we have simulated AFM spin waves under the impact of Gilbert damping and have fitted the data to damped oscillations. The decay length $\xi$ obtained from the fitted data is plotted against the wave number $k$ in Fig.\ \ref{fig:decay_length} where the circles mark the fitted data and the solid line shows the result of Eq.\ \eqref{eq:range_of_sw}.
	This illustrates the increasing complexity of spin wave driven AFM Skyrmions. For the Skyrmion motion due to a constant force we see in the simulations that the Gilbert damping acts on the Skyrmion like a friction force on a particle in classical mechanics. This coincides with the case of, e.g., current-driven AFM Skyrmions \cite{velkov}. 
	\begin{figure}[t!]
		\subfloat{\includegraphics[width=0.48\textwidth]{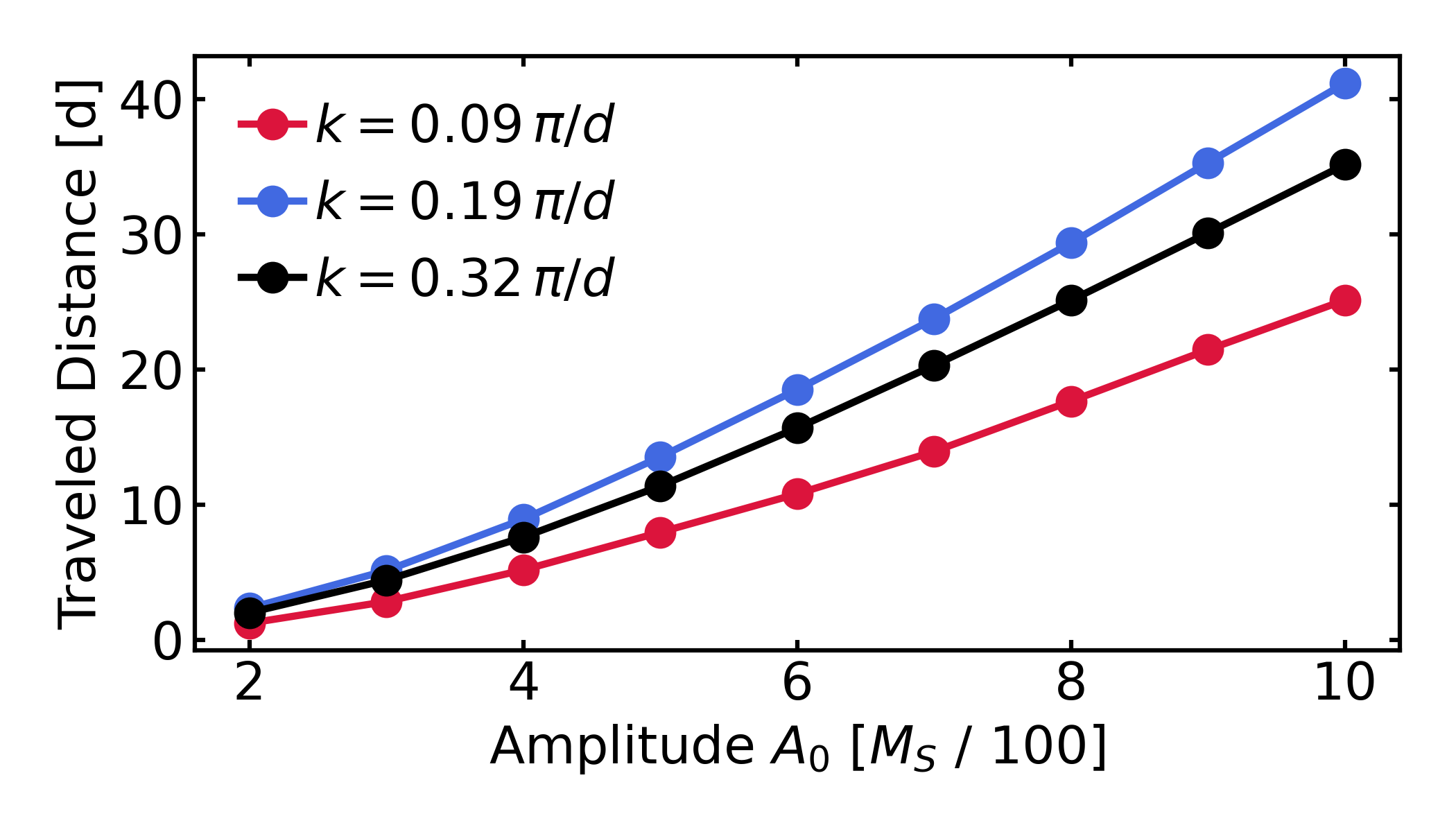}}
		\caption{(color online) Distance traveled by the Skyrmion after $t/t_0 = 600$ induced by a damped spin wave in dependence of its driving amplitude $A_0$ for the parameters $D/J=0.15$, $K/J=0.04$, $\alpha = 0.005$. Three different driving frequencies for the spin waves were considered.                                
			\label{fig:damping_range}}
	\end{figure}
	However, damped spin waves do not impose a constant force. Instead, Figs.\ \ref{fig:sk_acc_over_amp_circ} and \ref{fig:sk_acc_over_amp_lin} show that the acceleration of the Skyrmion strongly depends on the driving amplitude which decreases as the Skyrmion moves to the right, in the situation of continuous spin wave injection at the left edge. In conclusion, this results in a more involved Skyrmion motion which slows down to zero at some point. Hence, to accelerate Skyrmions in a damped system requires larger driving amplitudes. As an example, we show in Fig.\ \ref{fig:damping_range} how the spin-wave driven motion of a Skyrmion depends on the driving amplitude $A_0$ for $\alpha = 0.005$  for three different values of the spin wave numbers $k$. The Skyrmion starts $48$ lattice sites away from the spin wave source (left edge) and the traveled distance is taken after $t/t_0 = 600$ time steps. We see that spin waves with larger decay length $\xi$ are able to drive the Skyrmion further in space within a given time. This observation can be expected and illustrates that the driving amplitude plays a crucial role.
	Furthermore, the data in Fig.\ \ref{fig:decay_length} show that the spin waves are damped differently depending on their wave number. As a result, in systems with naturally occurring polychromatic spin waves, some parts of the wave are more likely to reach the Skyrmion. Overall, this will combine in a nonlinear manner the influences of spin wave amplitude and spin wave frequency of the Skyrmion motion. 
	
	\section{Application: Racetrack}
	\label{sec:racetrack}
\begin{figure}[!t]
		\begin{minipage}{0.24\textwidth}
			{\centering
			\includegraphics[width=.9\linewidth]{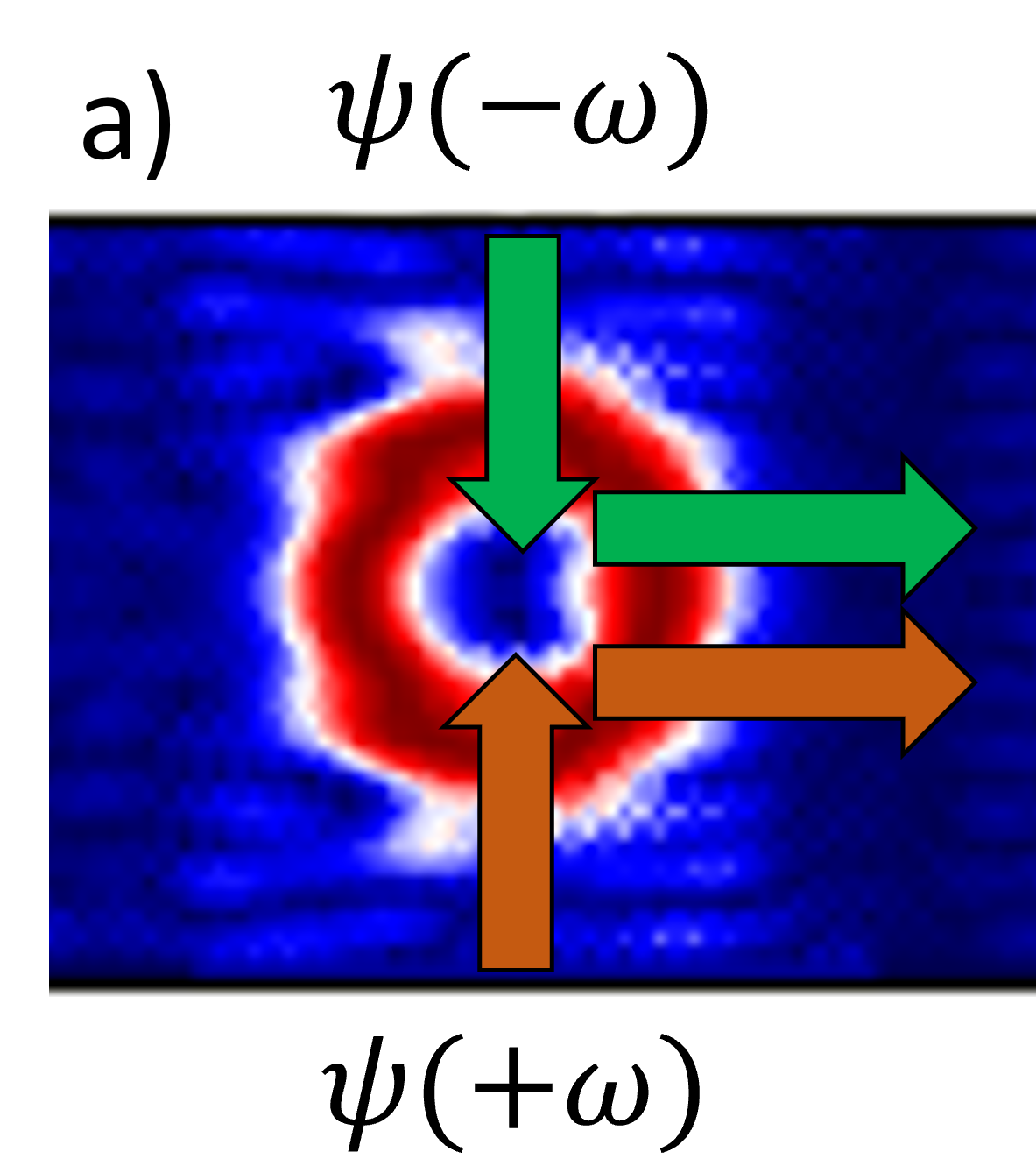}}
		\end{minipage}\hfill
		\begin{minipage}{0.24\textwidth}
			{\centering
			\includegraphics[width=.99\linewidth]{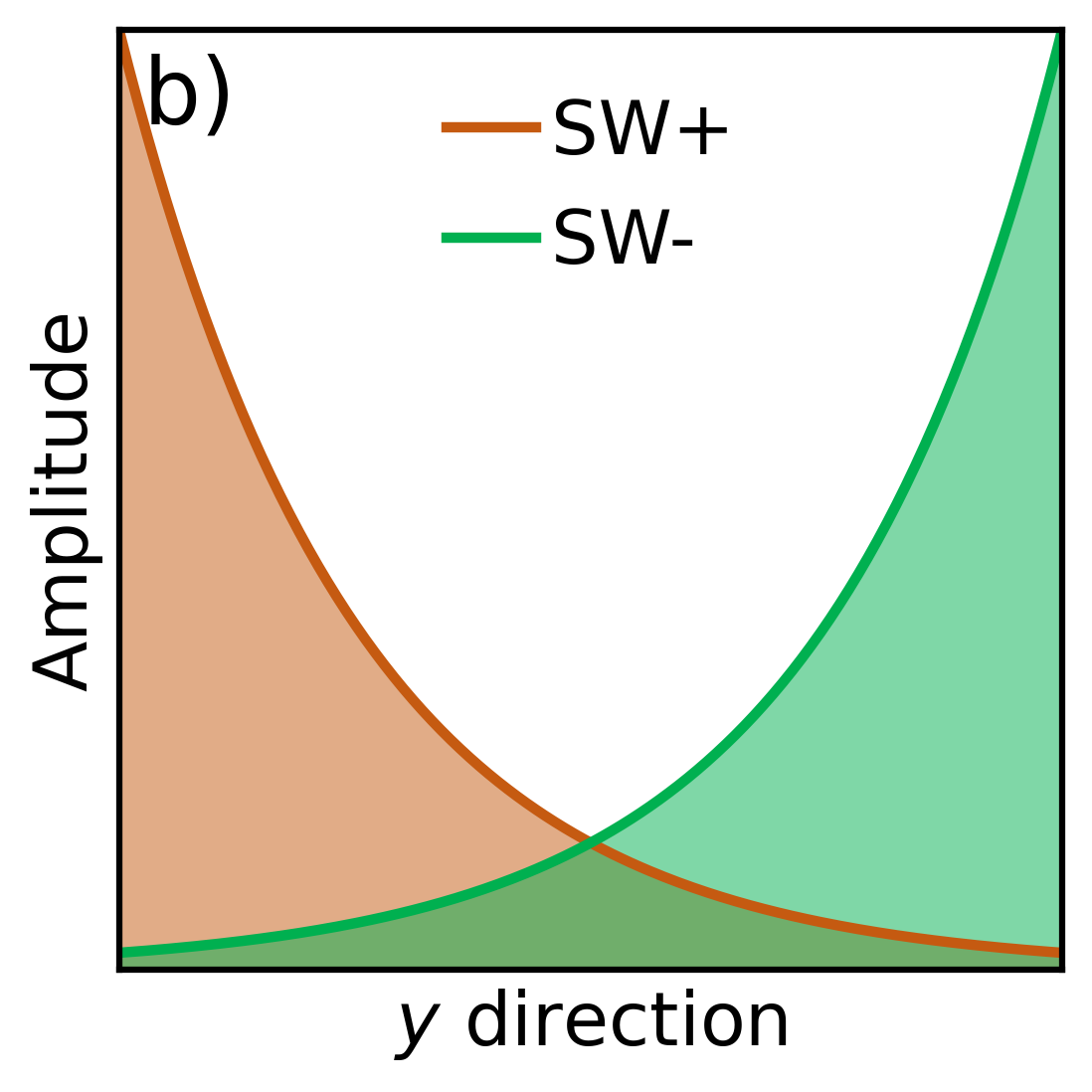}}
		\end{minipage}
		\caption{(color online) A possible scenario. (a) Skyrmion under the influence of the two spin waves from the top and bottom edge. The arrows indicate the direction of the forces the spin waves apply on the Skyrmion. The green arrows indicate the force induced by the clockwise rotating spin wave $\psi(-\omega)$ and the orange arrows indicate the force induced by the counterclockwise rotating spin wave $\psi(+\omega)$. (b) Amplitude of the two oppositely propagating spin waves which are damped over their course transverse to the racetrack.
			\label{fig:idea_racetrack}}
	\end{figure}

	In the following, we use the force, which magnons impose on Skyrmions in antiferromagnets, to propose an application in a racetrack. One popular set-up are narrow racetracks for Skyrmions \cite{fm_sk_racetrack, parkin2008}. However, spin waves injected at one end would quickly die out in amplitude with increasing distance from the injection point due to damping, as discussed in Sec.\ \ref{sec:damping}. Although AFM spin waves are, in general, able to move Skyrmions without a SHE (linearly polarized waves), we here propose a way to exploit the SHE using circularly polarized spin waves, injected as close as possible to the Skyrmion. To this end, we propose to apply circularly polarized spin waves to the racetrack from both sides, the upper and the lower, and use opposite chirality. The SHE induces forces with opposite $y$-components, but equal $x$-components, which ultimately drives the Skyrmion in the $x$-direction, along the track. More specifically, a left handed spin wave, injected from the upper side and propagating along the negative $y$-direction pushes the Skyrmion in the negative $y$-, and, due to the SHE, also in positive $x$-direction. On the lower edge sits a source which injects a right handed spin wave propagating along the positive $y$-direction. This wave pushes the Skyrmion in the positive $y$-direction, and also in the positive $x$-direction, see Fig.\ \ref{fig:idea_racetrack} (a). In total, the Skyrmion is thus pushed in the positive $x$-direction.
	As an advantage, the opposite spin wave forces in the $y$-direction create a potential which holds the Skyrmion in the center lane of the track: The farther away from the source the smaller is the amplitude of the spin wave (see Fig.\ \ref{fig:example_damped_wave}). Since the amplitude of the wave plays a crucial role for the strength of the force acting on the Skyrmion (see Fig.\ \ref{fig:sk_acc_over_amp_circ}), there exists a balanced $y$-coordinate in the middle of the racetrack where the forces in the positive and negative $y$-direction cancel each other. Then only the force in the $x$-direction remains from both the waves. This is sketched in Fig.\ \ref{fig:idea_racetrack} (b).
 
We have also explicitly simulated this concept for an antiferromagnetic spin-wave driven Skyrmion racetrack. Figure \ref{fig:racetrack_over_time} shows snapshots of the Skyrmion transport over time and in the Supplemental Material a video of this process can be found \cite{supp1}. 
	
	The Skyrmion will be pushed along the racetrack at each spin wave injection source, denoted by SW, and travels a small distance on its own due to inertia. When it reaches, even partially, the subsequent region of the driving spin waves it will be accelerated into the racetrack direction. We note that we consider a very narrow lattice since we use edge spin manipulation. This concept would also work on a broader lattice when it is possible to inject the spin waves anywhere. Then the racetrack will be determined by the injection source.
	\begin{figure}[t!]
		\subfloat{\includegraphics[width=0.50\textwidth]{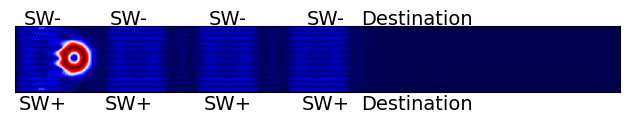}} \\
		\subfloat{\includegraphics[width=0.50\textwidth]{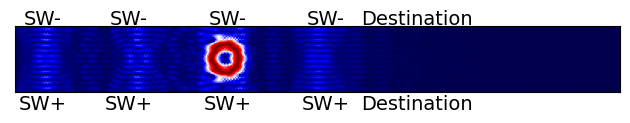}} \\
		\subfloat{\includegraphics[width=0.50\textwidth]{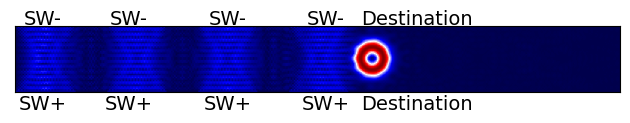}}
		\caption{(color online) Top to bottom: Position of the Skyrmion over time in our proposal of a racetrack. SW- and SW+ denote the injection source of left- and right handed circularly polarized spin waves, respectively, while the red circle is the Skyrmion. A video can be found in the Supplemental Material. 
			\label{fig:racetrack_over_time}}
	\end{figure}
	With regard to possible technical realization the transport along the racetrack does not depend on a particular spin wave frequency, but solely on the competition of the opposing circularly spin waves with opposite chirality. Thus, our concept of a racetrack is robust against technical imperfections or fluctuations of the spin waves. However, it comes with the price that the actual velocity of the Skyrmion is difficult to control.
	
	\section{Summary}
	
	We have shown that AFM spin waves scatter at Skyrmions and impose a net driving force on them, so that they start moving. AFM spin waves offer more flexibility in the control since, in contrast to the ferromagnetic counterpart, two linearly independent magnon modes exist. Circularly polarized spin waves create a Skyrmion Hall effect which non-trivially depends on the wave number $k$ and is symmetric with respect to the chirality of the spin wave. Linearly polarized spin waves, on the other hand, move the Skyrmion solely in the direction of the wave propagation. Rather, they drive Bloch and N{\'e}el Skyrmions differently with the symmetry that Bloch Skyrmions driven by $x$ polarized spin waves behave quantitatively and qualitatively exactly like N{\'e}el Skyrmions driven by $y$ polarized spin waves (same for a change of spin wave polarization).
	This is consistent with the fact that by a change of coordinates $x \rightarrow y$ and $y \rightarrow -x$ (rotation around $z$ axis) one can transform a N{\'e}el to a Bloch Skyrmion and can similarly transform an $x$-polarized spin wave to a $y$-polarized one as well. Both insights show that symmetries of the spin wave (sublattice symmetry for right- and left handed circularly polarized spin waves or component symmetry for linearly polarized ones) have an immediate impact on the force exerted on the Skyrmion. 

	In contrast to the ferromagnetic spin wave - Skyrmion interaction, AFM Skyrmions react much stronger to spin waves. Most importantly, they travel along their (longitudinal) propagation direction, and not against it as in the ferromagnetic case. 
	Our results show that probably most of the technical applications of the spin wave - Skyrmion interaction are indeed promising and work probably even better, with AFM Skyrmions. We propose one specific idea for illustrative purposes in this work.

	Furthermore, it is known that for ferromagnetic Skyrmions there is a competition between forces when temperature gradients are involved. They flow from cold to hot regions due to magnon forces or from hot to cold regions when entropy forces dominate \cite{sk_temp_1, sk_temp_2, sk_temp_3}. Thus, depending on which effect dominates, the ferromagnetic Skyrmion flows in a certain direction. This competition of forces is  not expected to exist for AFM Skyrmions since they flow with the magnon current. They should always move from hot to cold regions and should be more sensitive to temperature which would make them even more attractive for technical application in spin caloritronics. In fact, Ref.\ \cite{BarTre2016} reports that antiferromagnetic Skyrmions are much more sensitive to temperature. Also, a recent study found a non-monotonic temperature dependence of the velocities and the diffusion of current-driven skyrmion dynamics in ferrimagnets \cite{WeiNow2023}.

	Finally, we propose to use magnon waves to define a Skyrmion race track in bipartite antiferromagnets by injecting right-handed circularly polarized spin waves from the side along the track in $+y$-direction and at the same time left-handed circularly polarized spin waves in $-y$-direction. Due to ubiquitous damping, this channels Skyrmions of either type, N{\'e}el or Bloch, to move in $+x$-direction.
	
	\begin{acknowledgments}
		We thank Martin Stier for many fruitful discussions on Skyrmions. This work is funded by the Cluster of Excellence ''CUI: Advanced Imaging of Matter'' of the Deutsche Forschungsgemeinschaft (DFG) - EXC 2056 - project ID 390715994 (M.L. and M.T.). W.H. acknowledges support from the DFG SPP 2131 (Skyrmionics) - project ID 403505707.
	\end{acknowledgments}
	
	\appendix
	\section{ Damped Antiferromagnetic Spin Waves}
	\label{apendix1}
	The first few steps of the calculation are analogous to the case of the undamped AFM spin waves described in  Sec.\ \ref{sec:spinwave}. We linearize Eq.\ \eqref{eq:LLG} thereby considering a small but non-zero damping term $\alpha$, so that
	\begin{equation}
		\label{eq:lin_eom_damp}
		\begin{aligned}
			\delta \dot{\bm{M}}_i(t) = &-\delta \bm{M}_i (t) \times \bm{H}_i^\text{eff,0} - \bm{M}_i^0 \times \delta \bm{H}_i^\text{eff}(t)\\ 
			&+ \alpha \bm{M}_i^0 \times \delta \dot{\bm{M}}_i(t).   
		\end{aligned}
	\end{equation}
	Rearranging the terms we obtain
	\begin{equation}
		\label{eq:lin_eom_damp_final}
		\begin{aligned}
			(1 + \alpha^2) \delta \dot{\bm{M}}_i(t) = &-\delta \bm{M}_i (t) \times \bm{H}_i^\text{eff,0} - \bm{M}_i^0 \times \delta \bm{H}_i^\text{eff}(t)\\ 
			&- \alpha (\bm{M}_i^0 \cdot \bm{H}_i^\text{eff,0}) \delta \bm{M}_i(t) + \alpha \delta \bm{H}_i^\text{eff}(t).   
		\end{aligned}
	\end{equation}
	In this formalism the damping can be calculated over time where we assume that the  assumptions made in Sec.\ \ref{sec:spinwave} still hold and both sublattices experience the same damping $\alpha$. Next, we consider only terms linear in $\alpha$. Thus, the linearized equations of motion in the sublattice representation including damping are 
	\begin{align}
		\label{eq:lin_eom_subl_damp}
		\delta \dot{\bm{a}}_i &= -\delta \bm{a}_i \times \rho \bm{z} - C_{\bm{k}} \delta \bm{b}_{i} \times \bm{z} - \alpha \rho \delta \bm{a}_i - \alpha  C_{\bm{k}} \delta \bm{b}_{i}\, , \\
		\label{eq:lin_eom_subl_damp_b}
		\delta \dot{\bm{b}}_i &= \delta \bm{b}_i \times \rho \bm{z} + C_{\bm{k}} \delta \bm{a}_{i} \times \bm{z} - \alpha \rho \delta \bm{b}_i - \alpha  C_{\bm{k}} \delta \bm{a}_{i},
	\end{align}
	cf.\ Eqs.\ \eqref{eq:afm_lin_eom_new} and \eqref{eq:afm_lin_eom_new_ii}. 
	Again, we have assumed the sublattice spin waves to be circularly oscillating waves with different amplitudes. Since the spins are real valued, we assume, for non-vanishing damping the physical spin wave to be the real part of a complex vector $\delta \bm{b} = \Re(\phi_B)$ rotating in the complex plane
	\begin{equation}
		\label{eq:assume_deltaa}
		\phi_B = b \begin{pmatrix}
			1 \\
			i \\
			0
		\end{pmatrix} \exp(-i[\bm{k} \bm{r} - \omega t]),
	\end{equation}
	and similar for the other sublattice A. With these assumptions one can reduce Eqs.\ \eqref{eq:lin_eom_subl_damp} and \eqref{eq:lin_eom_subl_damp_b} to a system of equations regarding the sublattice wave amplitudes $a$ and $b$, which can be transformed to an eigenvalue equation
	\begin{equation}
		\label{eq:matrix_vec_damped}
		\omega 
		\begin{pmatrix}
			a \\
			b
		\end{pmatrix} = \begin{pmatrix}
			-\rho (1-i\alpha) & -C_{\bm{k}} (1-i\alpha) \\
			C_{\bm{k}} (1+i\alpha)& \rho (1+i\alpha)
		\end{pmatrix} \begin{pmatrix}
			a \\
			b
		\end{pmatrix},
	\end{equation}
	cf.\ Eq.\ \eqref{eq:matrix_vec}. The eigenvalues of the matrix in Eq.\ \eqref{eq:matrix_vec_damped} are
	\begin{equation}
		\label{eq:eigval_damped}
		\omega(\bm{k}) = i \alpha \rho \pm \sqrt{\rho^2 - C_{\bm{k}}^2},
	\end{equation}
	neglecting terms of order $\mathcal{O}(\alpha^2)$. This result matches expectations that the real part of the eigenvalue resembles the left- and right-handed oscillations of the non-damped case as in Eq.\ \eqref{eq:disp_rel} while the imaginary part describes time dependent damping $\exp(-\alpha \rho t)$ of the amplitudes. Analogous to Sec.\ \ref{sec:spinwave}, we may calculate the normalized eigenvectors which in the damped case are complex,
	\begin{equation}
		\label{eq:norm_eigvec_damp}
		\bm{\nu}(-\omega) = \begin{pmatrix}
			-\nu_2\\
			-\nu_1 (1 + i \alpha)
		\end{pmatrix}\; \text{  and  } \; 
		\bm{\nu}(\omega) = \begin{pmatrix}
			\nu_1\\
			\nu_2 (1 + i\alpha)
		\end{pmatrix},
	\end{equation}
	with $(\nu_1,\nu_2)$ as in Eq.\ \eqref{eq:eig_vec_exp}. The consequences of these complex entries become visible with regarding the previously made assumption in Eq.\ \eqref{eq:assume_deltaa}. For example consider a clockwise rotating spin wave on the sublattice B. By plugging in $\nu_2 (1 + i \alpha)$ for $b$, which is the second entry of $\bm{\nu}(\omega)$ from Eq.\ \eqref{eq:norm_eigvec_damp}, and regarding only the real part the wave reads
	\begin{equation}
		\label{eq:example_damped_da}
		\delta \bm{b}_i = \nu_2\begin{pmatrix}
			\cos(\bm{k} \bm{r}_i - \omega t) + \alpha \sin(\bm{k} \bm{r}_i - \omega t) \\
			\sin(\bm{k} \bm{r}_i - \omega t) - \alpha \cos(\bm{k} \bm{r}_i - \omega t)
		\end{pmatrix}  e^{-\alpha \rho t}.
	\end{equation}
	One can see that a $\pi$-phase shifted oscillation which is weighted by $\alpha$ is added to each component of the wave. This can be brought into a more convenient form by using the trigonometric properties $\cos(x) + \alpha \sin(x) = \sqrt{1+\alpha^2} \cos(x + \arctan(-\alpha))$ and $\sin(x) - \alpha \cos(x) = -\sqrt{1+\alpha^2} \cos(x + \arctan(1/\alpha))$, where here $x=\bm{k} \bm{r}_i - \omega t$. For small $\alpha$, we expand the inconvenient terms up to linear order in $\alpha$ and obtain for the spin waves with $\omega > 0$ the result
	\begin{equation}
		\label{eq:da_damped_omneg}
		\delta \bm{b}_i(\omega) = \nu_2 \begin{pmatrix}
			\cos(\bm{k} \bm{r}_i - \omega t - \alpha) \\
			\sin(\bm{k} \bm{r}_i - \omega t - \alpha)
		\end{pmatrix} e^{-\alpha \rho t}\, .
	\end{equation}
	This is a damped, clockwise rotating spin wave with a phase shift of $\alpha$ relative to $t=0$. 
	
	\section{Different Skyrmion Sizes}
\begin{figure}[t!]
		\subfloat{\includegraphics[width=0.48\textwidth]{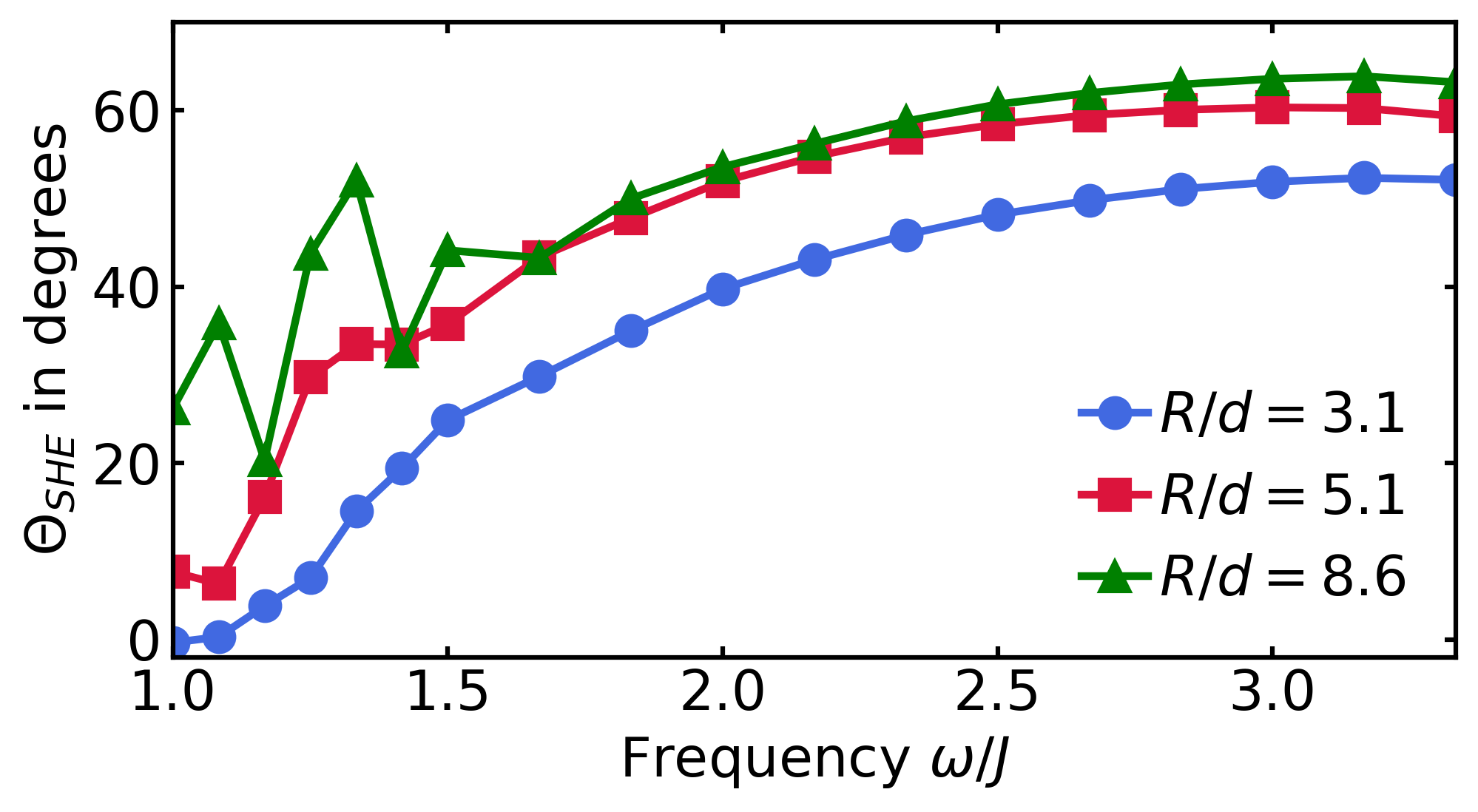}}
		\caption{(color online) Skyrmion Hall angle $\Theta_{\text{SHE}}$ for Skyrmions of different radii $R$ versus spin wave frequency $\omega$, shown in degrees. We have used the parameter $K/J=0.04$ and different values of $D/J$  to realize different Skyrmion radii.
			\label{fig:SM_she}}
	\end{figure}

	For the simulations of spin wave-Skyrmion interactions shown in the main text, we have used the system parameter $D/J=0.15$ and $K/J=0.04$. Thus, only Skyrmions with a radius $R/d=5.1$ were considered. The radius is calculated by
	\begin{equation}
		\label{eq:SM_rad}
		R = \frac{\sum_{x,y} q_{x,y}^2 \left[(x_{\text{Sk}}-x)^2 + (y_{\text{Sk}}-y)^2\right]}{\sum_{xy} q_{xy}^2},
	\end{equation}
	where the sum $\sum_{x,y}$ goes over the whole lattice and $q_{x,y}$ is the topological charge density $q_{x,y} = \bm{n}_{x,y} \cdot \left(\partial_x \bm{n}_{x,y} \times \partial_y \bm{n}_{x,y}\right)$. The vector $\bm{n}_{x,y}$ denotes here the vector on each lattice site, representing the magnetic moment, where each vector of sublattice B was formerly multiplied by $-1$.	
	To study Skyrmions of different sizes, we have performed simulations with two other DMI parameters $D$ resulting in different Skyrmion sizes, $D/J =  0.13 \rightarrow R/d = 3.1$ and $D/J =  0.17 \rightarrow R/d = 8.6$. One has to be careful when drawing conclusion between Skyrmion size and spin wave induced motion. Although the (linearized) spin waves are independent of the DMI when they are in the vicinity of the AFM ground state, this is most probably not the case when they scatter at the Skyrmion. Nonetheless, we simulated the wave number $k$ dependent acceleration of the Skyrmion for different Skyrmion sizes.
	\begin{figure}[t!]
		\subfloat{\includegraphics[width=0.48\textwidth]{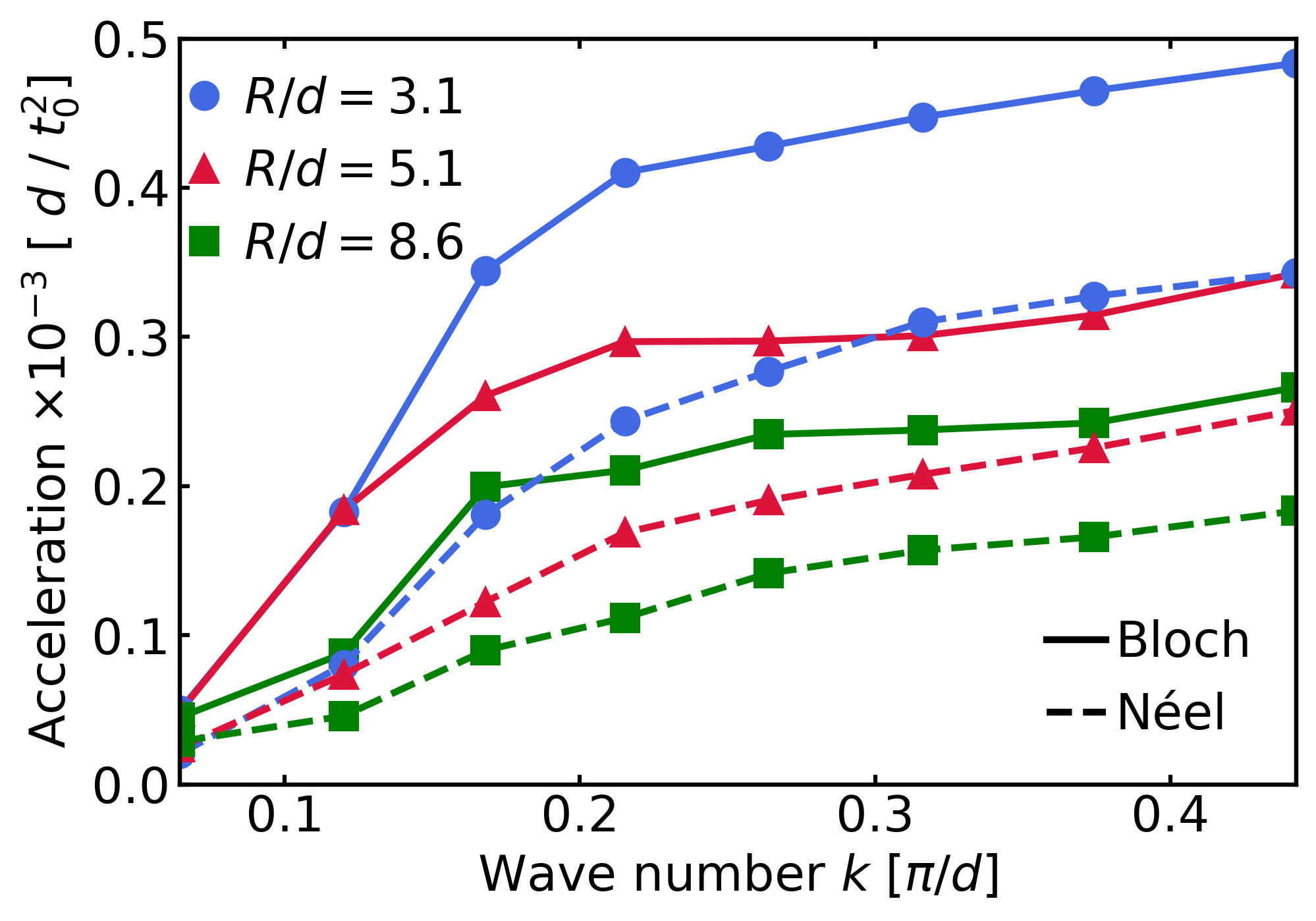}}
		\caption{(color online) Longitudinal acceleration of Bloch Skyrmions (solid lines) and N{\'e}el Skyrmions (dotted lines) due to  linearly polarized spin waves with different wave numbers $k$. The different symbols mark the different radii $R$ of the investigated Skyrmions. We have used the parameter $K/J=0.04$ and different values of $D/J$ in order to get different Skyrmion sizes.
			\label{fig:SM_acc_lin}}
	\end{figure}
	In Fig.\ \ref{fig:SM_acc_lin} the acceleration of Skyrmions impacted by $x$ linearly polarized spin waves with an amplitude of $A_0 = 0.02 M_S$ in dependence of the wave number $k$ is shown for different Skyrmion sizes and types. The solid lines indicate a Bloch Skyrmion while the dotted lines indicate a N{\'e}el-type Skyrmion. One can clearly see that for all sizes the Bloch-type Skyrmion is accelerated stronger than the N{\'e}el one. Additionally, smaller Skyrmions are accelerated stronger, however this may also be due to the lower DMI strength $D$. The size of the Skyrmion, or the DMI strength, also has an impact on the Skyrmion Hall angle regarding the Skyrmion motion induced by circularly polarized spin waves. In Fig.\ \ref{fig:SM_she}, the Skyrmion Hall angle in degree is plotted over the driving frequency $\omega$ of the spin wave for different Skyrmion sizes.
	Over the whole frequency range the SHE is more pronounced for larger Skyrmion sizes. While for smaller Skyrmions the Skyrmion Hall angle smoothly increases with increasing frequencies, it is sensitive to small frequency changes in the low-frequency regime for larger Skyrmions. Nevertheless, it is important to realize that in the low-frequency regime the absolute Skyrmion acceleration is low, too.
	
	\section{Comparison to Ferromagnetic Spin-Wave Driven Skyrmion}
\begin{figure}[t!]
		\subfloat{\includegraphics[width=0.48\textwidth]{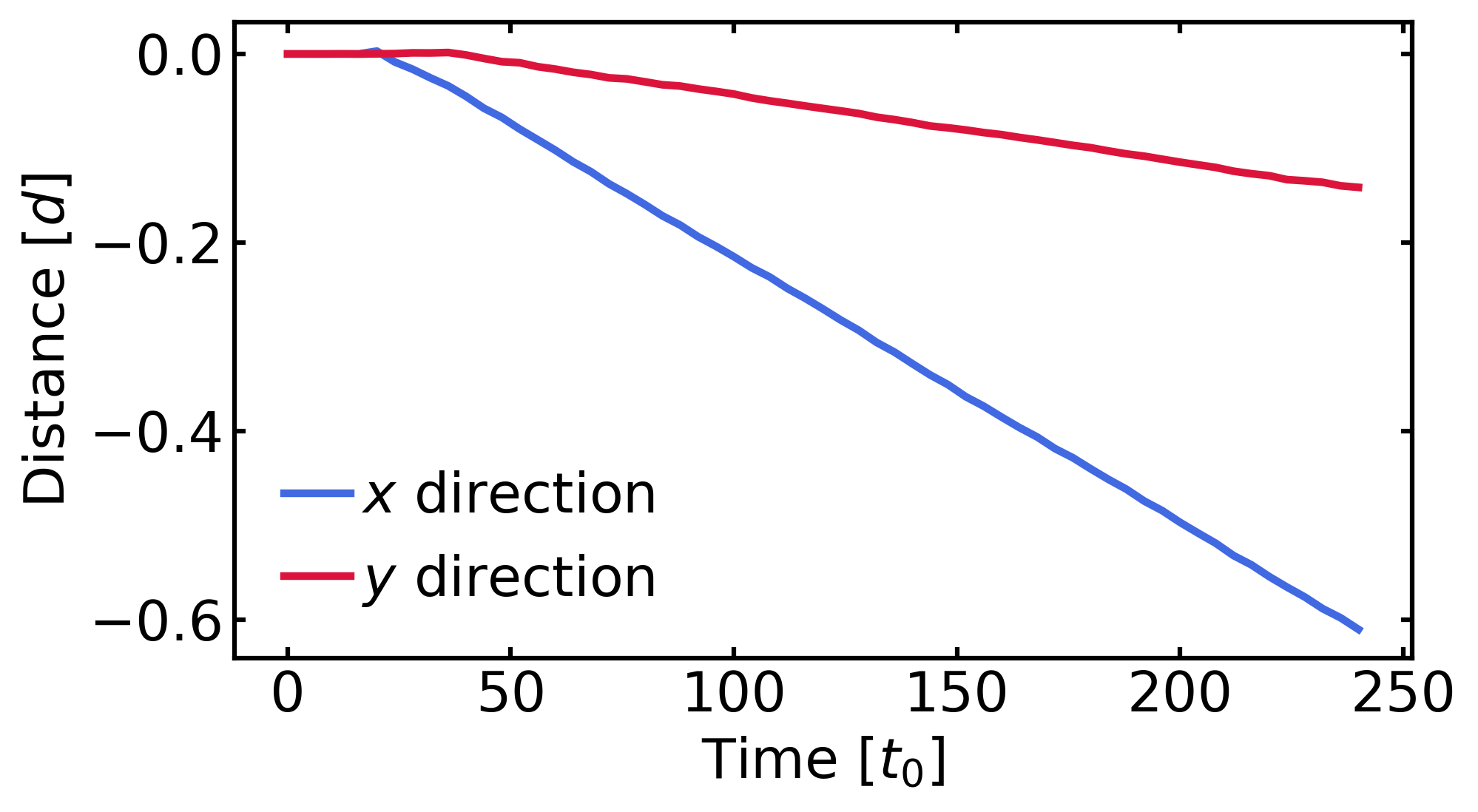}}
		\caption{(color online) Distance traveled  in $x$- and $y$ direction (blue and red, respectively) by a FM Skyrmion influenced by a monochromatic spin wave. The parameters are $K/J=0.04$ and $D/J = 0.15$ with the spin wave parameters  $\omega/J=1.8$  and amplitude  $A/M_S = 0.05$.
			\label{fig:SM_fm}}
	\end{figure}

	In this section, we compare the efficiency of driving Skyrmions by spin waves for AFM and FM Skyrmions.  We have simulated both FM and AFM Skyrmions within the same 2D lattice model and show in Fig.\ \ref{fig:SM_fm} the distance traveled by a FM Skyrmion in $x$- and $y$ direction (blue and red, respectively) as an example.
	The system parameter were chosen comparable to the simulations of the AFM case discussed in the main text. The spin wave frequency was fixed to an intermediate value in the spectrum  to $\omega/J = 1.8 \rightarrow k/d \approx 1.4$. Noticeably, the motion shows a constant velocity as it is typical for a FM Skyrmion. Although the AFM Skyrmions realize commonly an accelerated motion, one could compare Fig.\ \ref{fig:SM_fm} (FM) with Fig.\ \ref{fig:sk_motion_over_time_circ} of the main text (AFM). It turns out that the magnitude of the traveled distance in the AFM case, even in a fraction of the time, is much larger than the distance in the FM case.


\begin{thebibliography}{99}
\bibitem{footnote0} Both the exchange interaction and the DMI consists of interaction between direct neighboring spins. That means, this could be seen as sublattice interaction. Considering anti-parallel sublattices would only change the sign of these interactions.
\end{thebibliography}
\end{document}